\documentclass[aps,twocolumn]{revtex4}%
\usepackage{amsfonts}
\usepackage{amsmath}
\usepackage{amssymb}
\usepackage{graphicx}%
\providecommand{\U}[1]{\protect\rule{.1in}{.1in}}

\begin{document}
\preprint{HEP/123-qed}
\title[ ]{(An)Isotropic models in scalar and scalar-tensor cosmologies}
\author{Jos\'e Antonio Belinch\'on}
\affiliation{Departamento de F\'{\i}sica At\'{o}mica, Molecular y
Nuclear. \\ Universidad Complutense de Madrid, E-28040 Madrid,
Espa\~{n}a} \keywords{Scalar tensor cosmologies, Self-similar
solutions, Bianchi models} \pacs{PACS number}

\begin{abstract}
We study how the constants $G$ and $\Lambda$ may vary in different
theoretical models (general relativity with a perfect fluid,
scalar cosmological models (\textquotedblleft
quintessence\textquotedblright) with and without interacting
scalar and matter fields and a scalar-tensor model with a
dynamical $\Lambda$) in order to explain some observational
results. We apply the program outlined in section II to study
three different geometries which generalize the FRW ones, which
are Bianchi \textrm{V}, \textrm{VII}$_{0}$ and \textrm{IX}, under
the self-similarity hypothesis. We put special emphasis on
calculating exact power-law solutions which allow us to compare
the different models. In all the studied cases we arrive to the
conclusion that the solutions are isotropic and noninflationary
while the cosmological constant behaves as a positive decreasing
time function (in agreement with the current observations) and the
gravitational constant behaves as a growing time function.

\end{abstract}
\volumeyear{year}
\volumenumber{number}
\issuenumber{number}
\eid{identifier}
\date{\today}
\startpage{1}
\endpage{ }
\maketitle


\section{Introduction}

In a series of recent papers (\cite{Tony1}-\cite{Tony2}) we have studied and
compared how the "constants" $G$ and $\Lambda$ may vary in different
theoretical frameworks for several metrics. These different theoretical
frameworks are: general relativity with time varying constants (TCV), scalar
cosmological models with non-interacting scalar and matter fields and TVC, and
the last of the studied models is the usual scalar-tensor theory with a
dynamical cosmological constant which seems to be the most natural theoretical
model to study the possible variation of the gravitational and the
cosmological constants. In those recent works we have been able to state and
prove general results valid for all the geometries (all the Bianchi types as
well as for the FRW geometries) within the context of self-similar solutions
(SSS). We have focused our attention on this class of solutions since, as has
been pointed out by Coley (\cite{ColeyDS}),\emph{ }the self-similar models
play an important role in describing the asymptotic dynamics of the Bianchi
models. A large class of orthogonal spatially homogeneous models (including
all class B models) are asymptotically self-similar at the initial singularity
and are approximated by exact perfect fluid or vacuum self-similar power-law
models. In the same way, exact self-similar power-law models can also
approximate general Bianchi models at intermediate stages of their evolution
and which is also important, self-similar solutions can describe the behaviour
of Bianchi models at late times. Working under the hypothesis of
self-similarity allows us to find exact power-law solutions in such a way that
we may compare the obtained solution for each studied model.

In this paper we extend our program by considering the important case of the
interacting scalar and matter fields within the framework of scalar
cosmological models, and of course, we formulate the corresponding model with
TVC. Therefore the aim of the present work consists in studying and comparing,
by calculating the exact solution, how the constants $G$ and $\Lambda$ may
vary in different theoretical models and with different geometries under the
self-similar hypothesis. We apply the outlined program to study three
different geometries which generalize the FRW ones, which are Bianchi
\textrm{V}, \textrm{VII}$_{0}$ and \textrm{IX}, since as pointed out in
(\cite{Nilsson}) contrary to what is widely believed, an almost isotropic
cosmic microwave background (CMB) temperature does not imply that the universe
is \textquotedblleft close to a Friedmann-Lemaitre universe\textquotedblright%
\ and for this reason it is important to study anisotropic models.

For this purpose, in section II we start by summarizing the main results that
we employ in the next sections. In section IIA, we begin by setting the
notation and explaining how to calculate a homothetic vector field (HVF). We
also give the definitions of the some geometrical and kinematical quantities,
such as the anisotropic and the deceleration parameters. In section IIB we
outline the so called classical models. These models are the general
relativity one with a perfect fluid and its TVC model derived from it. We show
the general form of the solutions in each case. Section IIC is devoted to the
scalar cosmological models. We review the results stated in (\cite{Tony1})
showing the admitted form for the scalar field and its potential for every
homothetic solution (independently of the geometrical model). We study the
following cases:

\begin{itemize}
\item The standard scalar cosmological models with $G^{\prime}=0,$ i.e. where
the gravitational constant is considered as a true constant and its
corresponding generalization, i.e. a scalar model with $G^{\prime}\neq0.$

\item The non-interacting case of a scalar and matter fields and its TVC generalization.

\item As a part of our enlarging program, we now study the interesting case of
an interacting scalar and matter fields. We compare three \textquotedblleft a
priori\textquotedblright\ (in principle) different approaches. The first one
proposed by Maia et al (\cite{Lima}) where the authors do not split the
resulting conservation equation following a thermodynamical approach. The
second one has been proposed by Wetterich (\cite{Werterich}). In this case the
author consider the splitting conservation equations and takes as the possible
coupling between the scalar and matter field the function $q^{\phi}=\delta
\rho_{m},$ where the constant $\delta$ must be negative. The third of the
studied cases is the one proposed by Billard et al (\cite{Alan1}) where in
analogy with the model proposed by Wetterich but following a different point
of view the authors consider as a coupling function $Q=\delta H\rho_{m},$
where, in this case the constant $\delta$ must be positive in order to satisfy
the second law of thermodynamics.

\item We conclude the study of the scalar cosmological models by considering
the case of an interacting scalar and matter field with TVC. For this purpose
we split the generalized conservation equation and consider as a possible
coupling function $Q=\delta H\rho_{m},$ with $\delta>0.$
\end{itemize}

In section IID we consider a general scalar-tensor theory with a dynamical
cosmological constant, $\Lambda.$

Once we have presented the program, then we apply it, step by step, to three
different geometries which are Bianchi \textrm{V}, Bianchi \textrm{VII}$_{0}$
and Bianchi \textrm{IX}, since they may be considered as an anisotropic
generalization of the FRW geometries. In section III we study the Bianchi V
model, while in sections IV and V are studied the models Bianchi
\textrm{VII}$_{0}$ and \textrm{IX,} respectively. We begin each section by
deducing the metric through their Killing vector fields\ (KVF) and outlining
the Einstein tensor. Then we go next to study the classical, scalar and
scalar-tensor models as we have described previously. In each case we
calculate the standard solution, i.e. with the constants acting as true
constants and its TVC solution in order to show how such hypotheses modify the
standard solution. We end in section VI with a brief conclusions.

In the appendix A we have included the proofs of the results stated in section
II concerning the interacting scalar and matter models. We give a rigorous
proofs of the results by using the Lie group method. We also explore a
generalization which allows us to consider different coupling functions
between the scalar and the matter fields. In appendix B we prove the results
for the case of an interacting scalar and matter fields with TVC.

\section{The models}

\subsection{Self-similar solutions}

Throughout the paper, $\mathcal{M}$ will denote the usual smooth (connected,
Hausdorff, 4-dimensional) spacetime manifold with smooth Lorentz metric $g$ of
signature $(-,+,+,+)$ (see for example \cite{MC}). Thus $\mathcal{M}$ is
paracompact. A semi-colon and the symbol $\mathcal{L}$ denote the covariant
and Lie derivative, respectively. We shall use a system of units where $c=1$.
For a metric, $g,$ and for a vector field $\mathcal{H}\in\mathfrak{X}%
(\mathcal{M})$, $\left(  \mathcal{H}=\mathcal{H}_{i}(t,x,y,z)\partial_{x_{i}%
}\right)  _{i=1}^{4}$ the homothetic equation reads (\cite{Wainwrit} and
\cite{Carr}):%
\begin{equation}
\mathcal{L}_{\mathcal{H}}g=2g. \label{SS_Eq}%
\end{equation}
In the case of the Bianchi models, this equation brings us to obtain that the
scale factors behave as follows%
\begin{equation}
a=a_{0}\left(  t+t_{0}\right)  ^{a_{1}},\,\,b=b_{0}\left(  t+t_{0}\right)
^{a_{2}},\,\,d=d_{0}\left(  t+t_{0}\right)  ^{a_{3}}, \label{scafac}%
\end{equation}
where $\left(  a_{i}\right)  _{i=1}^{3}\in\mathbb{R}^{+}.$ In that follows we
define, $H=h\left(  t+t_{0}\right)  ^{-1},$ with $h=a_{1}+a_{2}+a_{3}.$ In
each of the studied cases we will get restrictions on the constants $\left(
a_{i}\right)  _{i=1}^{3}.$ We define the deceleration parameter as%
\begin{equation}
q=\frac{3}{h}-1, \label{q}%
\end{equation}
and the anisotropic quantities, $\mathcal{A}$ (\cite{H}-\cite{KB}), and
$\mathcal{W}^{2}$ (\cite{W}):%
\begin{equation}
\mathcal{A=}\frac{\sigma^{2}}{H^{2}},\qquad\mathcal{W}^{2}\mathcal{=}%
\frac{I_{3}}{H^{4}},
\end{equation}
that give us a measure of the anisotropy. $\sigma^{2}$ is the shear scalar and
$I_{3}=C^{abcd}C_{abcd},$ is the Weyl scalar.

\subsection{The classical models}

The field equations (FE) read%
\begin{equation}
R_{ij}-\frac{1}{2}Rg_{ij}=8\pi GT_{ij}-\Lambda g_{ij},\quad T_{i\,;j}^{j}=0,
\label{FE01}%
\end{equation}
where we consider vacuum solutions by making $T_{ij}=\Lambda=0,$ and the
classical perfect fluid solutions if $\Lambda=0.$ For a perfect fluid (PF) the
energy-momentum tensor is defined by:
\begin{equation}
T_{ij}^{m}=\left(  \rho+p\right)  u_{i}u_{j}+pg_{ij}, \label{eq00}%
\end{equation}
where $\rho$ is the energy density of the fluid, $p$ the pressure and they are
related by the equation of state $p=\omega\rho,$ $\left(  \omega
\in(-1,1]\right)  ,$ $u_{i}$ is the $4-$velocity. From the conservation
equation, $T_{i\,;j}^{j}=0,$ and taking into account that the scale factors
are given by Eq. (\ref{scafac}), then it is easy to arrive to the conclusion
that the energy density behaves as
\begin{equation}
\rho=\rho_{0}\left(  t+t_{0}\right)  ^{-2}. \label{enerden}%
\end{equation}

In order to take into account the variations of $G$ and $\Lambda$ we use the
Bianchi identities (see for example \cite{Lau}-\cite{Krori})
\begin{equation}
\left(  R_{ij}-\frac{1}{2}Rg_{ij}\right)  ^{;j}=\left(  8\pi GT_{ij}-\Lambda
g_{ij}\right)  ^{;j}, \label{eq8}%
\end{equation}
which read:%
\begin{equation}
8\pi G\left[  \rho^{\prime}+\rho\left(  1+\omega\right)  H\right]
=-\Lambda^{\prime}-8\pi G^{\prime}\rho, \label{laura3}%
\end{equation}
($^{\prime}$ means time derivative) in our case we obtain (assuming the
additional condition, $T_{i\,;j}^{j}=0):$%
\begin{equation}
\rho^{\prime}+\rho\left(  1+\omega\right)  H=0,\quad\text{and}\quad
\Lambda^{\prime}=-8\pi G^{\prime}\rho. \label{FE02}%
\end{equation}
It could be proven that the solution for this model (and for every Bianchi
model) has the following form%
\begin{align}
\rho &  =\rho_{0}\left(  t+t_{0}\right)  ^{-\alpha},\quad G=G_{0}\left(
t+t_{0}\right)  ^{\alpha-2},\nonumber\\
\Lambda &  =\Lambda_{0}\left(  t+t_{0}\right)  ^{-2}, \label{sol-tvc-1}%
\end{align}
where $\alpha=h\left(  \omega+1\right)  \in\mathbb{R},$ and $\rho_{0}%
,G_{0},\Lambda_{0}\in\mathbb{R}.$

\subsection{Scalar models}

We consider the following cases:

\textbf{1.-} For a scalar field $\phi$, the stress-energy tensor may be
written in the following form (\cite{Ellis}):%
\begin{equation}
T_{ij}^{\phi}=\left(  p_{\phi}+\rho_{\phi}\right)  u_{i}u_{j}+p_{\phi}g_{ij},
\label{scalar}%
\end{equation}
where the energy density and the pressure of the fluid due a scalar field are
given by%
\begin{equation}
\rho_{\phi}=\frac{1}{2}\phi^{\prime2}+V(\phi),\quad p_{\phi}=\frac{1}{2}%
\phi^{\prime2}-V(\phi), \label{defscal}%
\end{equation}
and conservation equation now reads (the Klein-Gordon equation (KG))%
\begin{equation}
\phi^{\prime\prime}+H\phi^{\prime}+\frac{d}{d\phi}V=0. \label{CE-S1}%
\end{equation}

In a previous work (\cite{Tony1}) we have proven that the unique form for the
scalar field, $\phi,$ and the potential, $V(\phi),$ compatible with the
self-similar solution (SSS) are given by%
\begin{equation}
\phi=\pm\sqrt{\alpha}\ln\left(  t+t_{0}\right)  ,\,\,V=\beta\exp\left(
\mp\frac{2}{\sqrt{\alpha}}\phi\right)  , \label{RS1}%
\end{equation}
where $\alpha\in\mathbb{R}^{+},$ while $\beta\in\mathbb{R}.$

\textbf{2.-} We would like to study how the gravitational constant varies when
we are considering a scalar field. For this purpose, by using the Bianchi
identity $\left(  G(t)T^{\phi}\right)  _{i\,;j}^{j}=0,$ we get%
\begin{equation}
\phi^{\prime}\left(  \phi^{\prime\prime}+H\phi^{\prime}+\frac{dV}{d\phi
}\right)  =-\frac{G^{\prime}}{G}\rho_{\phi}, \label{G-var-Scal-1}%
\end{equation}
which is the modified KG equation. In the $G-$var framework, in (\cite{Tony1})
we have proven that the main quantities behave as follows%
\begin{align}
\phi &  =\phi_{0}\left(  t+t_{0}\right)  ^{-\alpha},\quad V=\beta\left(
t+t_{0}\right)  ^{-2\left(  \alpha+1\right)  },\nonumber\\
G  &  =G_{0}\left(  t+t_{0}\right)  ^{g}, \label{RS2}%
\end{align}
where $\alpha,\phi_{0},G_{0},\beta\in\mathbb{R},$ and $g=2\alpha.$ Note that
they must verify $G\phi^{\prime2}\thickapprox t^{-2},$ $GV\thickapprox
t^{-2}.$ For a similar approach in the context of a FRW model with holographic
dark energy with varying gravitational constant see for example (\cite{Griego}).

Scalar and matter fields models. In this case the stress-energy tensor may be
defined by, $T=T^{m}+T^{\phi},$ where $T^{m}$ is defined by Eq. (\ref{eq00})
and $T^{\phi}$ by Eq. (\ref{scalar}). We may consider two cases: The
non-interacting and the interacting one (with their TVC versions).

\textbf{3.-} In the non-interacting case we already know how each quantity
must behave since we consider the conservation equations ($\operatorname{div}%
T^{m}=0=\operatorname{div}T^{\phi}$) so they behave as%
\begin{align}
\phi &  =\pm\sqrt{\alpha}\ln\left(  t+t_{0}\right)  ,\,\,V=\beta\exp\left(
\mp\frac{2}{\sqrt{\alpha}}\phi\right)  ,\nonumber\\
\rho &  =\rho_{0}\left(  t+t_{0}\right)  ^{-2}. \label{RS3}%
\end{align}
This class of solutions are known in the literature as scaling cosmological
solutions (\cite{Alan1} and the references therein).

\textbf{4.-} For the non-interacting case with $G-$var we consider, by taking
into account the Bianchi identity, the following modified KG equation (see
\cite{Tony1} for details)
\begin{equation}
G\tilde{\rho}^{\prime}+G\left(  \tilde{p}+\tilde{\rho}\right)  H=-G^{\prime
}\tilde{\rho},
\end{equation}
where $\tilde{\rho}=\rho_{m}+\rho_{\phi},$ and $\tilde{p}=p_{m}+p_{\phi},$
which we may split into%
\begin{align}
\rho_{m}^{\prime}+\left(  \rho_{m}+p_{m}\right)  \theta &  =-\frac{G^{\prime}%
}{G}\rho_{m},\label{NI1}\\
\phi^{\prime}\left(  \square\phi+\frac{dV}{d\phi}\right)   &  =-\frac
{G^{\prime}}{G}\rho_{\phi}, \label{NI2}%
\end{align}
where we already know how each quantity behaves, namely%
\begin{align}
\phi &  =\phi_{0}\left(  t+t_{0}\right)  ^{-\alpha},\quad V=\beta\left(
t+t_{0}\right)  ^{-2\left(  \alpha+1\right)  },\nonumber\\
G  &  =G_{0}\left(  t+t_{0}\right)  ^{2\alpha},\quad\rho=\rho_{0}\left(
t+t_{0}\right)  ^{-2\left(  \alpha+1\right)  }. \label{RS4}%
\end{align}

\textbf{5.-} In the case of an interacting perfect fluid with a scalar field,
interacting quintessence, we may consider several possibilities. In these
models, it is considered that the scalar field decays into the perfect fluid
and therefore it may be interpreted as a model with dark matter (DM) coupled
to dark energy (DE). In order to analyze this case we consider three different approaches:

\begin{enumerate}
\item \textbf{Approach 1}. (\cite{Lima}) In this case the conservation
equation reads%
\begin{equation}
\rho_{m}^{\prime}+\left(  \omega+1\right)  \rho_{m}H=-\phi^{\prime}\left(
\square\phi+\frac{dV}{d\phi}\right)  . \label{Lima}%
\end{equation}
In the appendix we will prove that the behaviour of the quantities are given
by Eq. (\ref{RS3}). In (\cite{Lima}) the authors interpret the scheme from a
thermodynamical point of view, a la Prigogine (\cite{Prigo}) i.e. a model with
matter creation.

\item \textbf{Approach 2}. (\cite{Werterich}) In this case the conservation
equation is split as follows%
\begin{align}
\rho_{m}^{\prime}+\left(  \omega+1\right)  \rho_{m}H  &  =-\phi^{\prime
}q^{\phi},\label{Wett0}\\
\square\phi+\frac{dV}{d\phi}  &  =q^{\phi}, \label{Wett}%
\end{align}
where $q^{\phi}=\delta\rho_{m}.$ Therefore we have obtained, see appendix,
that the quantities behave as Eq. (\ref{RS3}). Setting $\delta=0,$ we regain
the non-interacting case. The only constrain for the model is $\delta<0.$ The
author considers the possible coupling of the scalar field to matter from
previous works (\cite{wett2}).

\item \textbf{Approach 3}. (\cite{Alan1}) In this case the conservation
equations ($\left(  T^{m}\right)  _{i\,;j}^{j}=\delta H\rho_{m},$\ and
$\left(  T^{\phi}\right)  _{i\,;j}^{j}=-\delta H\rho_{m},$) read%
\begin{align}
\rho_{m}^{\prime}+\left(  \omega+1\right)  \rho_{m}H  &  =Q,\label{Al1}\\
\phi^{\prime}\left(  \square\phi+\frac{dV}{d\phi}\right)   &  =-Q, \label{Al2}%
\end{align}
where $Q=\delta H\rho_{m}.$ Again, in the appendix we will show that the
behaviour of the quantities are given by Eq. (\ref{RS3}). As above, the energy
of the scalar field is transferred to the matter field. The only restriction
is to assume $\delta>0,$ in such a way that the second law of thermodynamics
is verified (\cite{Pavon1}).\ Note that if $\delta=0,$ then the
non-interacting case is regained. The authors obtain the interacting term,
$\delta H\rho_{m},$ by taking into account the conformal transformation that
relates the scalar-tensor theory to the scalar one.
\end{enumerate}

In the appendix we will discuss the most general form of the function $Q,$ in
the framework of the self-similar solutions.

\textbf{6.-} To end, we consider the case of interacting fluids within the
framework of $G-$varying. For this model the modified KG equation reads%
\[
\rho_{m}^{\prime}+\phi^{\prime\prime}\phi^{\prime}+\frac{dV}{d\phi}%
\phi^{\prime}+H\left(  \phi^{\prime2}+\left(  \rho_{m}+p_{m}\right)  \right)
=
\]%
\begin{equation}
=-\frac{G^{\prime}}{G}\left(  \rho_{m}+\frac{1}{2}\phi^{\prime2}+V\right)  ,
\label{IG-var}%
\end{equation}
where we may split (as above) it in order to get%
\begin{align}
\rho_{m}^{\prime}+\left(  \rho_{m}+p_{m}\right)  H+\frac{G^{\prime}}{G}%
\rho_{m}  &  =Q,\label{ISG1}\\
\phi^{\prime}\left(  \phi^{\prime\prime}+\phi^{\prime}H+\frac{dV}{d\phi
}\right)  +\frac{G^{\prime}}{G}\rho_{\phi}  &  =-Q, \label{ISG2}%
\end{align}
with for example%
\begin{equation}
Q=\delta H\rho_{m}. \label{Q}%
\end{equation}

In this case, the solution takes the form (see the appendix for details):%
\begin{align}
\phi &  =\phi_{0}\left(  t+t_{0}\right)  ^{-\alpha},\;V(t)=\beta\left(
t+t_{0}\right)  ^{-2\left(  \alpha+1\right)  },\nonumber\\
G  &  =G_{0}\left(  t+t_{0}\right)  ^{2\alpha},\;\rho_{m}=\rho_{0}\left(
t+t_{0}\right)  ^{-2\left(  \alpha+1\right)  }. \label{RS6}%
\end{align}

For an alternative point of view see for example (\cite{Iranies}).

\subsection{Cosmological models with dynamical $\Lambda$ in scalar-tensor
theories}

We start with the action for the a general scalar-tensor theory of gravitation
with $\Lambda\left(  \phi\right)  ,$
\begin{align}
S  &  =\frac{1}{16\pi G_{\ast}}\int d^{4}x\sqrt{-g}\left[  \phi R-\frac
{\omega\left(  \phi\right)  g^{ij}\phi_{,i}\phi_{,j}}{\phi}+2\phi
\Lambda\left(  \phi\right)  \right] \nonumber\\
&  +S_{NG},
\end{align}
where $g=\det(g_{ij})$, $G_{\ast}$ is Newton's constant, $S_{NG}$ is the
action for the nongravitational matter. The arbitrary functions $\omega\left(
\phi\right)  $ and $\Lambda\left(  \phi\right)  $ distinguish the different
scalar-tensor theories of gravitation, $\Lambda\left(  \phi\right)  $ is a
potential function and plays the role of a cosmological constant, and
$\omega\left(  \phi\right)  $ is the coupling function of the particular
theory (\cite{will}).

The explicit field equations are%
\begin{align}
R_{ij}-\frac{1}{2}g_{ij}R  &  =\nonumber\\
&  \frac{8\pi}{\phi}T_{ij}+\Lambda\left(  \phi\right)  g_{ij}+\frac{\omega
}{\phi^{2}}\left(  \phi_{,i}\phi_{,j}-\frac{1}{2}g_{ij}\phi_{,l}\phi
^{,l}\right) \nonumber\\
&  +\frac{1}{\phi}\left(  \phi_{;ij}-g_{ij}\square\phi\right)  , \label{JBD1}%
\end{align}

\begin{equation}
\left(  3+2\omega\left(  \phi\right)  \right)  \square\phi=8\pi T-\frac
{d\omega}{d\phi}\phi_{,l}\phi^{,l}-2\phi\left(  \phi\frac{d\Lambda}{d\phi
}-\Lambda\left(  \phi\right)  \right)  , \label{JBD2}%
\end{equation}
where, $T=T_{i}^{i},$ is the trace of the stress-energy tensor, defined by Eq.
(\ref{eq00}). The gravitational coupling $G_{\mathrm{eff}}(t)$ is given by%
\begin{equation}
G_{\mathrm{eff}}(t)=\left(  \frac{2\omega+4}{2\omega+3}\right)  \frac{G_{\ast
}}{\phi(t)}. \label{G12}%
\end{equation}

In a recent work (see \cite{Tony2}) we have proven that the self-similar
solution admitted for the FE (\ref{JBD1}-\ref{JBD2}) have the following form%
\begin{equation}
\phi=\phi_{0}\left(  t+t_{0}\right)  ^{\tilde{n}},\,\,\,\Lambda\left(
\phi\right)  =\Lambda_{0}\phi^{\frac{-\left(  \tilde{n}+\alpha\right)
}{\tilde{n}}}, \label{JBDL}%
\end{equation}
with $\tilde{n}+\alpha=2,$ therefore $\Lambda\left(  t\right)  =\Lambda
_{0}\left(  t+t_{0}\right)  ^{-2}.$ The Brans-Dicke parameter is constant,
$\omega\left(  \phi\right)  =\mathrm{const}.,$ and $\rho=\rho_{0}\left(
t+t_{0}\right)  ^{-\alpha},$ $\alpha=\left(  1+\gamma\right)  h.$ Note that in
this case we have changed the notation and now the parameter of the EoS is
$\gamma,$ so $p=\gamma\rho.$

\section{Bianchi V models}

\subsection{The metric}

We consider (see for example \cite{MC}) the following Killing vectors fields
(KVF)%
\begin{equation}
\xi_{1}=\partial_{y},\quad\xi_{2}=\partial_{z},\quad\xi_{3}=\partial
_{x}+my\partial_{y}+mz\partial_{z}, \label{killingsBV}%
\end{equation}
where $m\in\mathbb{R}\backslash\left\{  0\right\}  ,$ such that, $\left[
\xi_{i},\xi_{j}\right]  =C_{ij}^{k}\xi_{k},$ i.e. $\left[  \xi_{1},\xi
_{2}\right]  =0,$ $\left[  \xi_{2},\xi_{3}\right]  =m\xi_{1},$ and $\left[
\xi_{1},\xi_{3}\right]  =m\xi_{1}.$ With these KVF we obtain the following
metric
\begin{equation}
ds^{2}=-dt^{2}+a(t)^{2}dx^{2}+b(t)^{2}e^{-2mx}dy^{2}+d(t)^{2}e^{-2mx}dz^{2}.
\label{BVm}%
\end{equation}

The Einstein tensor reads
\begin{align}
\frac{b^{\prime}a^{\prime}}{ba}+\frac{d^{\prime}a^{\prime}}{da}+\frac
{d^{\prime}b^{\prime}}{db}-\frac{3m^{2}}{a^{2}}  &  =G_{0}^{0},\label{fe1}\\
-2\frac{a^{\prime}}{a}+\frac{b^{\prime}}{b}+\frac{d^{\prime}}{d}  &
=G_{2}^{1}\label{fe2}\\
\frac{b^{\prime\prime}}{b}+\frac{d^{\prime\prime}}{d}+\frac{d^{\prime
}b^{\prime}}{db}-\frac{m^{2}c^{2}}{a^{2}}  &  =G_{1}^{1},\label{fe3}\\
\frac{d^{\prime\prime}}{d}+\frac{a^{\prime\prime}}{a}+\frac{d^{\prime
}a^{\prime}}{da}-\frac{m^{2}c^{2}}{a^{2}}  &  =G_{2}^{2},\label{fe4}\\
\frac{b^{\prime\prime}}{b}+\frac{a^{\prime\prime}}{a}+\frac{b^{\prime
}a^{\prime}}{ba}-\frac{m^{2}c^{2}}{a^{2}}  &  =G_{3}^{3}. \label{fe6}%
\end{align}

From Eq. (\ref{SS_Eq}) we obtain the following homothetic vector field%
\begin{equation}
\mathcal{H}=t\partial_{t}+\left(  1-a_{2}\right)  y\partial_{y}+\left(
1-a_{3}\right)  z\partial_{y}, \label{HOBV}%
\end{equation}
with the following constrains:%
\begin{equation}
a=a_{0}\left(  t+t_{0}\right)  ,\;b=b_{0}\left(  t+t_{0}\right)  ^{a_{2}%
},\;d=d_{0}\left(  t+t_{0}\right)  ^{a_{3}}, \label{restricBV}%
\end{equation}
where $a_{1}=1$ and $\left(  a_{i}\right)  _{i=2}^{3}\in\mathbb{R},$ and
therefore, $H=h\left(  t+t_{0}\right)  ^{-1},$ with $h=1+a_{2}+a_{3}.$ Note
that we have obtained a non-singular solution.

\subsection{Classical solutions}

\subsubsection{Vacuum solution}

The only possible solution is, $a_{2}=a_{3}=1,$ and $m=\pm1,$ hence the
solution is%
\begin{equation}
a=a_{0}\left(  t+t_{0}\right)  ,\,\,b=b_{0}\left(  t+t_{0}\right)
,\,\,d=d_{0}\left(  t+t_{0}\right)  , \label{vac_sol_BV}%
\end{equation}
with $m=\pm1.$ Therefore $q=0.$ With these results the metric yields%
\begin{equation}
ds^{2}=-dt^{2}+(t+t_{0})^{2}\left(  dx^{2}+e^{\pm2mx}\left(  dy^{2}%
+dz^{2}\right)  \right)  , \label{BVG4}%
\end{equation}
and therefore it admits more KVF, $\xi_{4}=-mz\partial_{y}+my\partial_{z},$
$\xi_{5}=-2mz\partial_{x}-2m^{2}yz\partial_{y}+\left(  m^{2}y^{2}-m^{2}%
z^{2}+e^{2mx}\right)  \partial_{z},$ and $\xi_{6}=2my\partial_{x}+\left(
m^{2}y^{2}-m^{2}z^{2}-e^{2mx}\right)  \partial_{y}+2m^{2}yz\partial_{z}.$ This
metric is the Milne form of a flat space-time (see Chapter 9, Eq. (9.8) of
\cite{WE}).

\subsubsection{Perfect fluid solution}

As we already know, the behaviour of the solution is given by Eq.
(\ref{enerden}), so it only remains to know the value of the constants
$\left(  a_{i}\right)  _{i=2}^{3}$ and $\omega,$ the parameter of the equation
of state. We have found the following solution for the FE\ (\ref{fe1}%
-\ref{fe6}) with the conservation equation, $\operatorname{div}T=0,$ where the
stress-energy tensor is defined by Eq. (\ref{eq00}),%
\begin{equation}
a_{1}=a_{2}=a_{3}=1,\;q=0,\;\omega=-\frac{1}{3},
\end{equation}
and $\forall m\in\left(  -1,1\right)  \backslash\left\{  0\right\}  .$ With
these results the metric collapses to Eq. (\ref{BVG4}), it does not inflate,
$q=0,$ and it is only valid for the EoS, $\omega=-\frac{1}{3},$ which is no
strange since for example the SSS for the Bianchi I model is only valid for
$\omega=1.$ A simple calculation shows us that the solution is isotropic since
the anisotropic parameters $\mathcal{A}$ and $\mathcal{W}^{2}$ vanish. From
the DS point of view, the solution is stable (\cite{WE}) since it is the FRW
model with negative curvature.

\subsubsection{Time varying constants model}

In this case the behaviour of the solution is given by Eq. (\ref{sol-tvc-1}).
The FE for this model are described by Eqs. (\ref{fe1}-\ref{fe6}) with the
conservation equations (\ref{FE02}), where $\alpha=\left(  \omega+1\right)  h$
and $h=\left(  1+a_{2}+a_{3}\right)  ,$ and the stress-energy tensor is
defined by Eq. (\ref{eq00}). Therefore we have found the next solution%
\begin{equation}
a_{1}=a_{2}=a_{3}=1,\;q=0,\;\alpha=3\left(  \omega+1\right)  ,
\end{equation}%
\[
G_{0}=\frac{\left(  1-m^{2}\right)  }{4\pi\rho_{0}\left(  \omega+1\right)
},\;\Lambda_{0}=3\left(  1-m^{2}\right)  -\frac{\left(  1-m^{2}\right)
}{\left(  \omega+1\right)  },
\]
with the following behaviour for $G$ and $\Lambda$%
\begin{align*}
G  &  \thickapprox\left\{
\begin{array}
[c]{l}%
\text{decreasing }\forall\omega\in\left(  -1,-1/3\right) \\
\text{constant if }\omega=-1/3\\
\text{growing }\forall\omega\in(-1/3,1]
\end{array}
\right.  ,\\
\Lambda_{0}  &  \thickapprox\left\{
\begin{array}
[c]{l}%
\text{negative }\forall\omega\in\left(  -1,-1/3\right) \\
\text{vanish if }\omega=-1/3\\
\text{positive }\forall\omega\in(-1/3,1]
\end{array}
\right.  ,
\end{align*}
this solution is valid $\forall\omega\in(-1,1],$ $m\in\left(  -1,1\right)
\backslash\left\{  0\right\}  .$ As above, with these results the metric
collapses to Eq. (\ref{BVG4}) and therefore it does not inflate. It is
isotropic and valid $\forall\omega\in(-1,1].$ As it is observed, if we set
$\omega=-1/3,$ then the solution collapses to the above one with $G=const.$
and $\Lambda=0.$ With regard to the behaviour of $G$ and $\Lambda,$ we only
may say that \textquotedblleft if\textquotedblright\ we take into account the
current observations (\cite{SNIa}) which suggest us that $\Lambda_{0}>0,$ then
the solution is only valid $\forall\omega\in(-1/3,1]$ which means that $G$ is
a growing time function, $G=G_{0}\left(  t+t_{0}\right)  ^{\alpha-2},$
$\alpha>2,$ $\forall\omega\in(-1/3,1]$.

\subsection{Scalar models}

\subsubsection{Scalar model}

The FE for this model are described by Eqs. (\ref{fe1}-\ref{fe6}) with the
stress-energy tensor is defined by Eqs. (\ref{scalar}-\ref{defscal}) and the
conservation equation given by Eq. (\ref{CE-S1}). Hence taking into account
Eq. (\ref{RS1}) we find the following result%
\begin{align}
a_{1}  &  =a_{2}=a_{3}=1,\quad q=0,\nonumber\\
\forall m  &  \in\left(  -1,1\right)  \backslash\left\{  0\right\}
,\;\alpha=\beta=2\left(  1-m^{2}\right)  .
\end{align}
With these results the metric collapses to Eq. (\ref{BVG4}), so it is not
inflationary and isotropic. As it has been proven in (see for example
\cite{KM} and for a extensive review of results \cite{ColeyDS} and all the
references therein) we cannot say anything about its dynamical behaviour since
the obtained value of $\alpha$ scape to this study. Nevertheless the study of
the solution through perturbations shows us that it is stable. The potential
behaves like the dynamical cosmological constant $V\sim\Lambda\sim t^{-2}.$

\subsubsection{Scalar model with G-var}

In this case, the model is described by the FE (\ref{fe1}-\ref{fe6}) with the
stress-energy tensor defined by Eq. (\ref{scalar}-\ref{defscal}) while the
conservation equations is defined by Eq. (\ref{G-var-Scal-1}). The solution
behaves as (\ref{RS2}). We find the following solution%
\begin{equation}
a_{1}=a_{2}=a_{3}=1,\quad q=0,
\end{equation}
$\forall m\in\left(  -1,1\right)  \backslash\left\{  0\right\}  ,$ therefore
the metric collapses to Eq. (\ref{BVG4}), so the solution is isotropic and not
inflationary, while%
\begin{equation}
\beta=\alpha^{2},\qquad G_{0}=\frac{2\left(  1-m^{2}\right)  }{\alpha^{2}}>0.
\end{equation}

With these results we do not know how $G\sim\left(  t+t_{0}\right)  ^{2\alpha
}$ may behave. Note that if $\alpha\in(-1,\infty)\backslash\left\{  0\right\}
,$ the potential, $V(t)\sim\left(  t+t_{0}\right)  ^{-2\left(  \alpha
+1\right)  },$ behaves as a decreasing time function, as a positive dynamical
cosmological constant, but $G$ is decreasing if $\alpha\in\left(  -1,0\right)
,$ and it behaves as a growing time function iff $\alpha>0$.

\subsubsection{Non-interacting scalar and matter model}

The geometric part of the FE for this model are given by Eqs. (\ref{fe1}%
-\ref{fe6}) with the stress-energy tensor defined by Eqs. (\ref{eq00} and
\ref{scalar}-\ref{defscal}) while the conservation equations are defined by
Eqs. ($\operatorname{div}T^{m}=0=\operatorname{div}T^{\phi}$). The solution
behaves as (\ref{RS3}). We find the following solution%
\begin{align}
a_{1}  &  =a_{2}=a_{3}=1,\quad q=0,\nonumber\\
\alpha &  =\beta,\qquad\rho_{0}=3\left(  1-m^{2}-\frac{\beta}{2}\right)  ,
\end{align}
as it is observed the metric collapses to Eq. (\ref{BVG4}), so the solution is
not inflationary but it is isotropic and only valid for $\omega=-\frac{1}{3}.$
We have found the restriction, $\beta<2\left(  1-m^{2}\right)  ,$ $\forall
m\in\left(  -1,1\right)  \backslash\left\{  0\right\}  .$ Therefore
$\alpha=\beta\in\left(  0,2\right)  .$

\subsubsection{Non-interacting scalar and matter fields with G-var}

The FE for the model are given by Eqs. (\ref{fe1}-\ref{fe6}). The
stress-energy tensor is defined by Eqs. (\ref{eq00} and \ref{scalar}%
-\ref{defscal}) while the conservation equations are defined by Eqs
(\ref{NI1}-\ref{NI2}). The solution behaves as (\ref{RS4}), therefore we have
obtained the following solution%
\[
a_{1}=a_{2}=a_{3}=1,\quad q=0,\quad\forall m\in\left(  -1,1\right)
\backslash\left\{  0\right\}  ,
\]
so the metric collapses to Eq. (\ref{BVG4}), i.e. the solution is not
inflationary and isotropic and it is only valid for $\omega=-\frac{1}{3}.$ The
rest of the parameters are%
\begin{equation}
\beta=\alpha^{2},\qquad\rho_{0}=\frac{3}{G_{0}}\left(  1-m^{2}\right)
-3\alpha^{2},
\end{equation}
so, in order to get $\rho_{0}>0,$ we have the following restriction,
$\alpha^{2}<\frac{\left(  1-m^{2}\right)  }{G_{0}}.$ Unfortunately with these
results we do not know how $G\sim\left(  t+t_{0}\right)  ^{2\alpha}$ may
behave. Only by \textquotedblleft assuming\textquotedblright\ that the
potential, $V(t)\sim\left(  t+t_{0}\right)  ^{-2\left(  \alpha+1\right)  },$
mimics the dynamic of $\Lambda$ we may bound the value of $\alpha,$ that is,
$\alpha\in(-1,\infty)\backslash\left\{  0\right\}  ,$ and therefore, as above,
we may say that $G,$ is decreasing if $\alpha\in\left(  -1,0\right)  ,$ and it
behaves as a growing time function iff $\alpha>0$.

\subsubsection{Interacting scalar and matter model}

\paragraph{Approach 1.}

The FE are given by Eq. (\ref{fe1}-\ref{fe6}) and the stress-energy tensor
defined by Eqs. (\ref{eq00} and \ref{scalar}-\ref{defscal}) The conservation
equation by Eq. (\ref{Lima}). Since the behaviour of the quantities are given
by Eq. (\ref{RS3}), then we have found the following solution%
\begin{align}
a_{1}  &  =a_{2}=a_{3}=1,\quad q=0,\nonumber\\
\alpha &  =2\left(  1-m^{2}\right)  -\rho_{0}\left(  \omega+1\right)
,\nonumber\\
\beta &  =2\left(  1-m^{2}\right)  +\frac{\rho_{0}}{2}\left(  \omega-1\right)
,
\end{align}
$\forall m\in\left(  -1,1\right)  \backslash\left\{  0\right\}  ,$ and
$\ \forall\omega\in(-1,1],$ with the restriction $\rho_{0}<\frac{2\left(
1-m^{2}\right)  }{\left(  \omega+1\right)  }.$ With these results the metric
collapses to Eq. (\ref{BVG4}), therefore the solution does not inflate and it
is isotropic. For example, if $\omega=0,$ (which may stands for a mixture of
dark matter and dark energy) then%
\[
\beta=1-m^{2}+\frac{\alpha}{2},\quad\rho_{0}=2\left(  1-m^{2}\right)
-\alpha,
\]
with $\alpha<2\left(  1-m^{2}\right)  \in\left(  0,2\right)  .$

\paragraph{Approach 2.}

We have found the following solution for the FE (\ref{fe1}-\ref{fe6}) with the
stress-energy tensor is defined by Eqs. (\ref{eq00} and \ref{scalar}%
-\ref{defscal}) and the conservation equations given by Eqs. (\ref{Wett0}%
-\ref{Wett}). The behaviour of the quantities is given by Eq. (\ref{RS3}), so%
\begin{align}
a_{1}  &  =a_{2}=a_{3}=1,\quad q=0,\nonumber\\
\beta &  =2\left(  1-m^{2}\right)  -\frac{\left(  1-\omega\right)  }{2\left(
\omega+1\right)  }\left(  2\left(  1-m^{2}\right)  -\alpha\right)
,\nonumber\\
\rho_{0}  &  =\frac{2\left(  1-m^{2}\right)  -\alpha}{\omega+1},\quad
\alpha<2\left(  1-m^{2}\right)  ,\nonumber\\
\delta &  =-\frac{1}{\sqrt{\alpha}}\left(  3\omega+1\right)  ,
\end{align}
where $\omega>-1/3,$ and $\forall m\in\left(  -1,1\right)  \backslash\left\{
0\right\}  ,$ otherwise $\delta\geq0$ (remember that in this case $\delta$
must be negative). With these results the metric collapses to Eq. (\ref{BVG4})
so it is isotropic and non-inflationary. \ For example if $\ $we set
$\omega=0,$ then the solution reduces to:%
\[
\beta=1-m^{2}+\frac{\alpha}{2},\;\rho_{0}=2\left(  1-m^{2}\right)
-\alpha,\;\delta=-\frac{1}{\sqrt{\alpha}}.
\]
with $\alpha<2\left(  1-m^{2}\right)  .$

\paragraph{Approach 3.}

The FE (\ref{fe1}-\ref{fe6}) with the stress-energy tensor is defined by Eqs.
(\ref{eq00} and \ref{scalar}-\ref{defscal}) while the conservation equations
are given by Eqs.(\ref{Al1}-\ref{Al2}). As in the above studied cases the
behaviour of the quantities is given by Eq. (\ref{RS3}), so we get the
following solution%
\begin{align}
a_{1}  &  =a_{2}=a_{3}=1,\quad q=0,\nonumber\\
\rho_{0}  &  =\frac{2\left(  1-m^{2}\right)  -\alpha}{\omega+1},\qquad
\alpha<2\left(  1-m^{2}\right)  ,\nonumber\\
\beta &  =2\left(  1-m^{2}\right)  -\frac{\left(  1-\omega\right)  }{2\left(
\omega+1\right)  }\left(  2\left(  1-m^{2}\right)  -\alpha\right)
,\nonumber\\
\delta &  =\frac{1}{3}+\omega,\qquad\rho=\rho_{0}\left(  t+t_{0}\right)
^{-2},
\end{align}
and therefore the solution is isotropic and non-inflationary since the metric
collapses to Eq. (\ref{BVG4}). As it is observed, we have obtained the same
solution as in the approach 2. In this model, there is a critical value for
$\omega$, $\omega=-1/3,$ thus $\omega>-1/3,$ and $\forall m\in\left(
-1,1\right)  \backslash\left\{  0\right\}  ,$ otherwise $\delta\leq0$
(remember that in this case $\delta$ must be positive)$.$ As in the above
studies cases if we set $\omega=0$ then we get%
\[
\beta=2\left(  1-m^{2}\right)  +\frac{\alpha}{2},\;\rho_{0}=2\left(
1-m^{2}\right)  -\alpha,\;\delta=\frac{1}{3},
\]
with $\alpha<2\left(  1-m^{2}\right)  .$

As it is observed for $\omega=0$ we have obtained the same results in the
three approaches. The main difference with regard to the non-interacting case
is that in this case the solution is valid for $\forall\omega\in
(-1/3,1]\mathbf{,}$ except in the approach 1 where $\omega\in(-1,1].$

\subsubsection{Interacting scalar and matter fields with G-var}

The FE for this model are given by Eqs. (\ref{fe1}-\ref{fe6}). The
stress-energy tensor is defined by Eqs. (\ref{eq00} and \ref{scalar}%
-\ref{defscal}) while the conservation equations are given by Eqs.(\ref{ISG1}%
-\ref{ISG2}). In this case the behaviour of the quantities are given by Eq.
(\ref{RS6}), so we get the following solution with $Q=\delta H\rho_{m},$%
\[
a_{1}=a_{2}=a_{3}=1,\quad q=0,
\]
with $\forall m\in\left(  -1,1\right)  \backslash\left\{  0\right\}  ,$ so the
solution is non-inflationary and isotropic. The rest of parameters are%
\begin{align}
\omega &  =\frac{1}{G_{0}\rho_{0}}\left[  2\left(  1-m^{2}\right)
-G_{0}\left(  \alpha^{2}+\rho_{0}\right)  \right]  ,\nonumber\\
\beta &  =\frac{1}{2G_{0}}\left[  6\left(  1-m^{2}\right)  -G_{0}\left(
\alpha^{2}+2\rho_{0}\right)  \right]  ,\nonumber\\
\delta &  =\frac{1}{3G_{0}\rho_{0}}\left[  6\left(  1-m^{2}\right)
-G_{0}\left(  3\alpha^{2}+2\rho_{0}\right)  \right]  ,
\end{align}
therefore the solution is valid for $\forall\omega\in(-1,1].$

If we set $\omega=0,$ then we obtain:%
\begin{align*}
\rho_{0}  &  =\frac{1}{G_{0}}\left[  2\left(  1-m^{2}\right)  -G_{0}\alpha
^{2}\right]  ,\\
\beta &  =\frac{1}{2G_{0}}\left[  2\left(  1-m^{2}\right)  +G_{0}\alpha
^{2}\right]  ,\qquad\delta=\frac{1}{3},
\end{align*}
so we have the following restriction, $\alpha^{2}<\frac{2\left(
1-m^{2}\right)  }{G_{0}},$ since $\rho_{0}>0.$ With these results we do not
know how $G\sim\left(  t+t_{0}\right)  ^{2\alpha}$ may behave$.$ Note that if
$\alpha\in(-1,\infty)\backslash\left\{  0\right\}  ,$ the potential,
$V(t)\sim\left(  t+t_{0}\right)  ^{-2\left(  \alpha+1\right)  },$ behaves as a
decreasing time function, as a positive dynamical cosmological constant, but
$G$ is decreasing if $\alpha\in\left(  -1,0\right)  ,$ and it behaves as a
growing time function iff $\alpha>0$.

\subsection{Scalar tensor model}

In this case the effective stress-energy tensor takes the following form%
\begin{align}
T_{0}^{0}  &  =\frac{8\pi}{\phi}\rho-H\frac{\phi^{\prime}}{\phi}+\frac{\omega
}{2}\left(  \frac{\phi^{\prime}}{\phi}\right)  ^{2}+\Lambda\left(
\phi\right)  ,\label{T1}\\
T_{1}^{1}  &  =-\frac{8\pi}{\phi}p-\frac{\phi^{\prime}}{\phi}\left(
\frac{d^{\prime}}{d}+\frac{b^{\prime}}{b}\right)  -\frac{\omega}{2}\left(
\frac{\phi^{\prime}}{\phi}\right)  ^{2}-\frac{\phi^{\prime\prime}}{\phi
}+\Lambda\left(  \phi\right) \\
T_{2}^{2}  &  =-\frac{8\pi}{\phi}p-\frac{\phi^{\prime}}{\phi}\left(
\frac{d^{\prime}}{d}+\frac{a^{\prime}}{a}\right)  -\frac{\omega}{2}\left(
\frac{\phi^{\prime}}{\phi}\right)  ^{2}-\frac{\phi^{\prime\prime}}{\phi
}+\Lambda\left(  \phi\right) \\
T_{3}^{3}  &  =-\frac{8\pi}{\phi}p-\frac{\phi^{\prime}}{\phi}\left(
\frac{a^{\prime}}{a}+\frac{b^{\prime}}{b}\right)  -\frac{\omega}{2}\left(
\frac{\phi^{\prime}}{\phi}\right)  ^{2}-\frac{\phi^{\prime\prime}}{\phi
}+\Lambda\left(  \phi\right)  \label{T4}%
\end{align}
and conservation equations%
\begin{align}
\left(  3+2\omega\left(  \phi\right)  \right)  \left(  \frac{\phi
^{\prime\prime}}{\phi}+H\frac{\phi^{\prime}}{\phi}\right)  -2\left(
\Lambda-\phi\frac{d\Lambda}{d\phi}\right)   &  =\frac{8\pi}{\phi}\left(
\rho-3p\right)  ,\label{T5}\\
\rho^{\prime}+\left(  \rho+p\right)  H  &  =0. \label{T6}%
\end{align}

For the FE (\ref{JBD1}-\ref{JBD2}) and taking into account Eq. (\ref{JBDL}) we
find the next solution.%
\begin{align}
a_{1}  &  =a_{2}=a_{3}=1,\;\phi_{0}=1,\;m=m,\qquad\tilde{n}=-1-3\gamma
,\nonumber\\
\rho_{0}  &  =-\frac{1}{8\pi\left(  1+\gamma\right)  }\left(  \left(
\omega+1\right)  \left(  3\gamma+1\right)  ^{2}+6\gamma+2m^{2}\right)
,\nonumber\\
\Lambda_{0}  &  =-\frac{\left(  \left(  \left(  \gamma-1\right)  \left(
3\gamma+1\right)  ^{2}\right)  \omega+2\left(  3\gamma+1\right)  \left(
m^{2}-1\right)  \right)  }{2\left(  1+\gamma\right)  },
\end{align}
therefore, the metric collapses to Eq. (\ref{BVG4}), so the solution is
isotropic an non inflationary. $\rho_{0}=0,$ iff%
\begin{align*}
\gamma &  =-\frac{\left(  \omega\pm\sqrt{2\omega\left(  1-m^{2}\right)
+3-2m^{2}}+2\right)  }{3\left(  \omega+1\right)  },\;\omega\neq-1,\\
\gamma &  =-\frac{1}{3}m^{2},\qquad\omega=-1,
\end{align*}
while, $\tilde{n}=0\Longleftrightarrow\gamma_{c}=-\frac{1}{3}.$ If
$\gamma<-1/3$ then $\tilde{n}>0$ and $\tilde{n}<0$ $\forall\gamma\in(-1/3,1].$

For example, if we set $\omega=3300$ (\cite{BDparameter}) and $m=\pm1/2,$ then
$\rho_{0}>0\Longleftrightarrow\gamma\in I_{1}$, where $I_{1}=\left(
-0.34054,-0.32633\right)  ,$ but if $m=\pm50,$ then $\rho_{0}<0,$
$\forall\gamma\in(-1,1],$ and if $m=\pm0.001,$ $\rho_{0}>0\Longleftrightarrow
\gamma\in I_{0},$ with $I_{0}=\left(  -0.34164,-0.32523\right)  .$ While
$\Lambda_{0}>0,$ $\forall\gamma\in(-1,1]\backslash I_{2},$ where
$I_{2}=\left(  -0.333446,-0.333333\right)  .$ As it is observed $\Lambda
_{0}=0,$ if
\[
\gamma=-\frac{1}{3},\quad\gamma=\frac{1}{3\omega}\left(  \omega\pm
\sqrt{2\omega\left(  2\omega-3m^{2}+3\right)  }\right)  ,
\]
note that $\Lambda_{0}<0,$ $\forall\gamma\in I_{2}.$

Therefore this solution has only physical meaning $\forall\gamma\in I_{1},$
i.e. a small neighborhood of $\gamma_{c},$ $\mathcal{E}(\gamma_{c}=-\frac
{1}{3})=I_{1},$ note that $I_{2}\subset I_{1}.$ For example for $\gamma_{c},$
then $\rho_{0}>0$ but $\Lambda_{0}=0$ and $\tilde{n}=0,$ therefore $\phi
=\phi_{0}$ and this means that $G=G_{0}$ (i.e. constant), i.e. for this
specific value, the solution is the same than the one obtained in the
classical model for a perfect fluid. In the same way we may compare the JBD
solution with the one obtained in the classical situation when $G$ and
$\Lambda$ vary. If $\gamma>\gamma_{c},$ $\forall\gamma\in I_{1}$ then
$\tilde{n}<0$ and therefore $G_{\mathrm{eff}}\thickapprox\phi^{-1},$ is a
growing time function, and $\Lambda$ behaves as a positive decreasing time
function. Note that $\tilde{n}>0,$ $\forall\gamma\in(-0.34164,\gamma_{c}),$
$\tilde{n}=0$ if $\gamma=\gamma_{c},$ and $\tilde{n}<0$ $\forall\gamma
\in(\gamma_{c},-0.32633).$

\section{Bianchi $\mathrm{VII}_{0}$}

\subsection{The metric}

We have the following Killing vector fields (\cite{MC}):
\begin{equation}
\xi_{1}=\partial_{y},\quad\xi_{2}=\partial_{z},\quad\xi_{3}=\partial
_{x}-z\partial_{y}+y\partial_{z}, \label{killings}%
\end{equation}
with $\left[  \xi_{1},\xi_{2}\right]  =0,$ $\left[  \xi_{2},\xi_{3}\right]
=-\xi_{1},$ $\left[  \xi_{1},\xi_{3}\right]  =\xi_{2},$so we may consider that
$C_{13}^{2}=1,C_{23}^{1}=-1,$ and therefore we have the following metric%
\begin{align}
ds^{2}  &  =-dt^{2}+a^{2}dx^{2}+\left(  b^{2}\cos^{2}x+d^{2}\sin^{2}x\right)
dy^{2}+\nonumber\\
&  2\cos x\sin x(b^{2}-d^{2})dydz+\left(  b^{2}\sin^{2}x+d^{2}\cos
^{2}x\right)  dz^{2}. \label{mBVII}%
\end{align}

The homothetic vector field is obtained from Eq. (\ref{SS_Eq}) is%
\begin{equation}
\mathcal{H}=\left(  t+t_{0}\right)  \partial_{t}+\left(  1-a_{2}\right)
y\partial_{y}+\left(  1-a_{2}\right)  z\partial_{z}, \label{HOBVII}%
\end{equation}
so the scale factors must behave as%
\begin{equation}
a=a_{0}\left(  t+t_{0}\right)  ,\;b=b_{0}\left(  t+t_{0}\right)  ^{a_{2}%
},\;d=d_{0}\left(  t+t_{0}\right)  ^{a_{2}}, \label{scaleBVII}%
\end{equation}
with $a_{1}=1,a_{2}=a_{3}\in\mathbb{R}^{+}.$ We emphasize the fact that we
have been able to obtain non-singular scale factors. So the restrictions are:%
\begin{equation}
a_{1}=1,\qquad a_{2}=a_{3}, \label{restricBVII}%
\end{equation}
hence the metric collapses to the following one%
\begin{equation}
ds^{2}=-dt^{2}+\left(  t+t_{0}\right)  ^{2}dx^{2}+\left(  t+t_{0}\right)
^{2a_{2}}\left(  dy^{2}+dz^{2}\right)  ,
\end{equation}
thus it belongs to a localy rotationally symmetric (LRS) $\mathrm{BVII}_{0}$
which looks like a LRS \textrm{BI}$\mathrm{.}$ In this way we find a new KVF,
$\xi_{4}=\partial_{x}.$

The FE read:%
\begin{align}
\frac{a^{\prime}}{a}\frac{b^{\prime}}{b}+\frac{a^{\prime}}{a}\frac{d^{\prime}%
}{d}+\frac{b^{\prime}}{b}\frac{d^{\prime}}{d}+\frac{1}{4a^{2}}\left(
2-\frac{d^{2}}{b^{2}}-\frac{b^{2}}{d^{2}}\right)   &  =8\pi GT_{1}%
^{1},\label{feb71}\\
\frac{b^{\prime\prime}}{b}+\frac{d^{\prime\prime}}{d}+\frac{b^{\prime}}%
{b}\frac{d^{\prime}}{d}+\frac{1}{4a^{2}}\left(  -2+\frac{d^{2}}{b^{2}}%
+\frac{b^{2}}{d^{2}}\right)   &  =-8\pi GT_{2}^{2},\\
\frac{b^{\prime\prime}}{b}-\frac{d^{\prime\prime}}{d}+\frac{a^{\prime}}%
{a}\frac{b^{\prime}}{b}-\frac{a^{\prime}}{a}\frac{d^{\prime}}{d}+\frac
{1}{a^{2}}\left(  \frac{b^{2}}{d^{2}}-\frac{d^{2}}{b^{2}}\right)   &  =0,\\
\frac{a^{\prime\prime}}{a}+\frac{b^{\prime\prime}}{b}+\frac{a^{\prime}}%
{a}\frac{b^{\prime}}{b}+\frac{1}{4a^{2}}\left(  2+\frac{b^{2}}{d^{2}}%
-\frac{3d^{2}}{b^{2}}\right)   &  =-8\pi GT_{3}^{3},\\
\frac{a^{\prime\prime}}{a}+\frac{d^{\prime\prime}}{d}+\frac{a^{\prime}}%
{a}\frac{d^{\prime}}{d}+\frac{1}{4a^{2}}\left(  2+\frac{d^{2}}{b^{2}}%
-\frac{3b^{2}}{d^{2}}\right)   &  =-8\pi GT_{4}^{4}, \label{feb75}%
\end{align}
and the conservation equation(s) for the different studied cases.

\subsection{Classical solutions}

\subsubsection{Vacuum solution}

There is a solution with $a_{2}=0.$ Therefore the metric collapses to this one%
\begin{equation}
ds^{2}=-dt^{2}+\left(  t+t_{0}\right)  ^{2}dx^{2}+dy^{2}+dz^{2}.
\end{equation}
This metric is the Taub form of a flat space-time (see chaper 9 Eq. (9.6) of
\cite{WE} and \cite{HW}).

\subsubsection{Perfect fluid solution}

The FE\ \ are given by Eqs. (\ref{feb71}-\ref{feb75}) with the conservation
equation, $\operatorname{div}T=0.$ The stress-energy tensor is defined by Eq.
(\ref{eq00}), \ and the general behaviour for the solution is given by Eq.
(\ref{enerden}). We find only two solutions

\begin{enumerate}
\item This solution is only valid if $\;\omega=1,$%
\begin{equation}
ds^{2}=-dt^{2}+\left(  t+t_{0}\right)  ^{2}dx^{2}+dy^{2}+dz^{2},
\end{equation}
which is the same metric as the obtained one in the vacuum solution.

\item
\begin{equation}
ds^{2}=-dt^{2}+\left(  t+t_{0}\right)  ^{2}\left(  dx^{2}+dy^{2}%
+dz^{2}\right)  , \label{FRWb7}%
\end{equation}
i.e. $a_{1}=a_{2}=a_{3}=1,$ $q=0,$ a flat FRW and only valid for
$\omega=-\frac{1}{3}.$ A simple calculation shows us the obtained solution is
non-inflationary, $q=0,$ and isotropic since the anisotropic parameters
$\left(  \mathcal{A},\mathcal{W}^{2}\right)  $ vanish.
\end{enumerate}

\subsubsection{Time-varying constants scene}

In this case the behaviour of the solution is given by Eq. (\ref{sol-tvc-1}).
The FE (\ref{feb71}-\ref{feb75}) with the conservation equations (\ref{FE02}),
where $\alpha=\left(  \omega+1\right)  h$ and $h=\left(  1+2a_{2}\right)  $.
and the stress-energy tensor is defined by Eq. (\ref{eq00}). Therefore we have
found the next solution by taking into account the obtained SS restrictions
for the scale factors given by eq. (\ref{restricBVII}). $G$ behaves as
follows:
\begin{equation}
G=G_{0}\left(  t+t_{0}\right)  ^{\alpha-2},\;G_{0}=\frac{A}{4\pi\rho
_{0}\left(  \omega+1\right)  \alpha}, \label{G}%
\end{equation}
where $A=2a_{2}+a_{2}^{2},$ while the cosmological \textquotedblleft
constant\textquotedblright\ behaves as:
\[
\Lambda=A\left(  1-\frac{2}{\alpha}\right)  \left(  t+t_{0}\right)
^{-2}=\Lambda_{0}\left(  t+t_{0}\right)  ^{-2}.
\]

We find the following solution
\begin{align}
a_{1}  &  =a_{2}=a_{3}=1,\;q=0,\quad\rho=\rho_{0}\left(  t+t_{0}\right)
^{-3\left(  \omega+1\right)  },\nonumber\\
G  &  =G_{0}\left(  t+t_{0}\right)  ^{3\left(  \omega+1\right)  -2}%
,\quad\Lambda=\Lambda_{0}\left(  t+t_{0}\right)  ^{-2},
\end{align}
valid $\forall\omega\in(-1,1],$ and therefore the metric collapses to Eq.
(\ref{FRWb7}). We find the following behaviour for $G$ and $\Lambda$%
\begin{align*}
G  &  \thickapprox\left\{
\begin{array}
[c]{l}%
\text{decreasing }\forall\omega\in\lbrack-1,-1/3)\\
\text{constant if }\omega=-1/3\\
\text{growing }\forall\omega\in(-1/3,1]
\end{array}
\right. \\
\Lambda_{0}  &  \thickapprox\left\{
\begin{array}
[c]{l}%
\text{negative }\forall\omega\in\lbrack-1,-1/3)\\
\text{vanish if }\omega=-1/3\\
\text{positive }\forall\omega\in(-1/3,1]
\end{array}
\right.  .
\end{align*}
As it is observed the solution is valid $\forall\omega\in(-1,1],$ instead of
only for $\omega=-1/3.$ For this value of EoS the solution collapses to the PF
solution, i.e. $G=const.$ and $\Lambda=0.$ The metric collapses to the given
one by Eq. (\ref{FRWb7}) so it is non-inflationary and isotropic. As in the
case of the \textrm{BV} solution, only under the assumption of $\Lambda>0$
(\cite{SNIa})$,$ we may say that $G$ behaves as a growing time function.

\subsection{Scalar models}

\subsubsection{Scalar model}

The FE are given by Eqs. (\ref{feb71}-\ref{feb75}) while the stress-energy
tensor is defined by Eqs. (\ref{scalar}-\ref{defscal}). For this model the
conservation equation is given by Eq. (\ref{CE-S1}). Hence, by taking into
account Eq. (\ref{RS1}) we find the following results

\begin{enumerate}
\item Vacuum solution
\begin{equation}
a_{1}=1,\quad a_{2}=a_{3}=0,\quad\alpha=\beta=0,
\end{equation}
therefore we have obtained the same solution as in the vacuum case.

\item A non-trivial solution
\begin{equation}
a_{1}=a_{2}=a_{3}=1,q=0,\quad\alpha=\beta=2.
\end{equation}
therefore for this solution the metric collapses to Eq. (\ref{FRWb7}) i.e. the
solution is non-inflationary and isotropic. From the DS point of view the
solution is stable (\cite{ColeyDS}).
\end{enumerate}

\subsubsection{Scalar model with G-var}

The FE of this model are given by Eqs. (\ref{feb71}-\ref{feb75}) with the
stress-energy tensor is defined by Eq. (\ref{scalar}-\ref{defscal}) while the
following conservation equations is defined by Eq. (\ref{G-var-Scal-1}). The
solution behaves as
\begin{align*}
\phi &  =\phi_{0}\left(  t+t_{0}\right)  ^{-\alpha},\quad V(t)=\beta\left(
t+t_{0}\right)  ^{-2\left(  \alpha+1\right)  },\\
G  &  =G_{0}\left(  t+t_{0}\right)  ^{2\alpha}.
\end{align*}
We find the following solution%
\begin{align}
a_{1}  &  =a_{2}=a_{3}=1,\quad q=0,\nonumber\\
\alpha &  =1,\quad\beta=1,\quad G_{0}=2.
\end{align}
As in the previous studied models, the metric collapses to Eq. (\ref{FRWb7}),
so it is non-inflationary and isotropic. As it is observed, since $\alpha=1,$
then the potential $V$ is positive and behaves as $V\sim t^{-4},$ while $G\sim
t^{2},$ i.e. $G$ is a growing time function. Note that $GV\sim t^{-2},$ while
in the standard approach $\Lambda\sim t^{-2}.$ Note that in the classical
approach $\Lambda$ is not multiplied by $G$ while in the quintessence approach
the potential $V,$ is multiplied by $G.$

\subsubsection{Non-interacting scalar and matter model}

The model is described by the following FE (\ref{feb71}-\ref{feb75}) with the
stress-energy tensor is defined by Eqs. (\ref{eq00} and \ref{scalar}%
-\ref{defscal}) while the following conservation equations is defined by Eqs.
($\operatorname{div}T^{m}=0=\operatorname{div}T^{\phi}$). The solution behaves
as (\ref{RS3}). We find the following solution%
\begin{align}
a_{1}  &  =a_{2}=a_{3}=1,\quad q=0,\nonumber\\
\alpha &  =\beta=2\left(  1-\frac{\rho_{0}}{3}\right)  ,\;\rho_{0}\in\left(
0,3\right)  ,\;\omega=-\frac{1}{3}.
\end{align}
With these results the metric collapses to Eq. (\ref{FRWb7}) and therefore it
is non-inflationary and isotropic. The solution is only valid for
$\omega=-\frac{1}{3},$ with the restriction $\rho_{0}\in\left(  0,3\right)  $
so if $\rho_{0}\ll1$ then the solution collapses to the one obtained in the
scalar model with $\alpha=\beta=2$.

\subsubsection{Non-interacting scalar and matter fields with G-var}

The FE are given by Eqs. (\ref{feb71}-\ref{feb75}) with the stress-energy
tensor defined by Eqs. (\ref{eq00} and \ref{scalar}-\ref{defscal}) and the
following conservation equations given by Eqs (\ref{NI1}-\ref{NI2}). The
solution behaves as (\ref{RS4}), therefore we have obtained the following
solution%
\begin{align}
a_{1}  &  =a_{2}=a_{3}=1,\quad q=0,\nonumber\\
\alpha &  =1=\beta,\;\rho_{0}=\frac{3}{G_{0}}-\frac{3}{2},\;\omega=-\frac
{1}{3}.
\end{align}
In order to get, $\rho_{0}>0,$ we have the following restriction: $G_{0}<2.$
The metric collapses to Eq. (\ref{FRWb7}) and therefore it is non-inflationary
and isotropic and only valid for $\omega=-\frac{1}{3}$. As in the scalar
solution with $G-$var, since $\alpha=1,$ then the potential $V$ is positive
and behaves as $V\sim t^{-4},$ while $G\sim t^{2},$ i.e. $G$ is growing time function.

\subsubsection{Interacting scalar and matter model}

\paragraph{Approach 1}

The model is described by Eqs. (\ref{feb71}-\ref{feb75}) where the
stress-energy tensor is defined by Eqs. (\ref{eq00} and \ref{scalar}%
-\ref{defscal}) and the following conservation equation (\ref{Lima}). Since
the behaviour of the quantities are given by Eq. (\ref{RS3}), then we have
found the following solution%
\begin{align}
a_{1}  &  =a_{2}=a_{3}=1,\quad q=0,\nonumber\\
\alpha &  =2-\rho_{0}\left(  \omega+1\right)  ,\;\beta=2+\frac{1}{2}\rho
_{0}\left(  \omega-1\right)  ,
\end{align}
this solution is valid $\forall\omega\in(-1,1]$, instead of a unique value of
$\omega$ as in the case of non-interacting fluids. We find only a restriction,
$\rho_{0}<\frac{2}{\left(  \omega+1\right)  }.$ The metric collapses to Eq.
(\ref{FRWb7}) and therefore it is isotropic and non-inflationary. For example
if we set $\omega=0,$ then we get%
\[
\alpha=2\left(  \beta-1\right)  ,\;\rho_{0}=2\left(  1-\beta\right)
,\;\forall\beta\in\left(  1,2\right)  .
\]

\paragraph{Approach 2}

We obtain the following solution for the FE (\ref{feb71}-\ref{feb75}) with
stress-energy tensor is defined by Eqs. (\ref{eq00} and \ref{scalar}%
-\ref{defscal}) and the following conservation equations given by Eqs.
(\ref{Wett0}-\ref{Wett}). The behaviour of the quantities are given by Eq.
(\ref{RS3}),
\begin{align}
a_{1}  &  =a_{2}=a_{3}=1,q=0,\nonumber\\
\alpha &  =2-\rho_{0}\left(  \omega+1\right)  ,\quad\beta=2+\frac{1}{2}%
\rho_{0}\left(  \omega-1\right)  ,\nonumber\\
\delta &  =-\frac{1+3\omega}{\sqrt{2-\rho_{0}\left(  \omega+1\right)  }},
\end{align}
valid $\forall\omega\in(-1/3,1],$ $\left(  \delta\leq0\right)  ,$ with the
same restriction, $\rho_{0}<\frac{2}{\left(  \omega+1\right)  }.$ The metric
collapses to Eq. (\ref{FRWb7}) so it is isotropic and non-inflationary. \ For
example if we set $\omega=0,$ then we get%
\[
\delta=-\frac{1}{\sqrt{\alpha}},\quad\beta=1+\frac{\alpha}{2},\quad\rho
_{0}=2-\alpha,
\]
$\forall\alpha\in\left(  0,2\right)  .$ Observe that we get the same result as
in the previous case.

\paragraph{Approach 3}

In this case the FE are given by Eqs. (\ref{feb71}-\ref{feb75}) while the
stress-energy tensor is defined by Eqs. (\ref{eq00} and \ref{scalar}%
-\ref{defscal}) and the conservation equations are given by Eqs.(\ref{Al1}%
-\ref{Al2}). As in the above studied cases the behaviour of the quantities are
given by Eq. (\ref{RS3}), so we get the following solution%
\begin{align}
a_{1}  &  =a_{2}=a_{3}=1,q=0,\nonumber\\
\alpha &  =2-\rho_{0}\left(  \omega+1\right)  ,\;\delta=\frac{1}{3}%
+\omega,\nonumber\\
\beta &  =2+\frac{1}{2}\rho_{0}\left(  \omega-1\right)  ,
\end{align}
valid $\forall\omega\in(-1/3,1],$ $\left(  \delta\geq0\right)  $with the
restriction, $\rho_{0}<\frac{2}{\left(  \omega+1\right)  }.$ The metric
collapses to Eq. (\ref{FRWb7}) and therefore it is non-inflationary and
isotropic. For example if we set $\omega=0,$ then we get%
\[
\delta=\frac{1}{3},\;\beta=1+\frac{\alpha}{2},\rho_{0}=2-\alpha,
\]
$\forall\alpha\in\left(  0,2\right)  .$

As it is observed we have obtained the same solution in the three cases. The
main difference with the non-interacting case is that for these models the
solution is valid for all values of $\omega,$ i.e. $\forall\omega\in(-1/3,1],$
which is more realistic. In approach 1 we have obtained $\omega\in(-1,1].$

\subsubsection{Interacting scalar and matter fields with G-var}

The FE (\ref{feb71}-\ref{feb75}) with the stress-energy tensor is defined by
Eqs. (\ref{eq00} and \ref{scalar}-\ref{defscal}) while the conservation
equations are given by Eqs.(\ref{ISG1}-\ref{ISG2}). In this case the behaviour
of the quantities are given by Eq. (\ref{RS6}), so we get the following
solution with $Q=\delta H\rho_{m},$%
\begin{align}
a_{1}  &  =a_{2}=a_{3}=1,\quad q=0,\nonumber\\
\omega &  =\frac{1}{G_{0}\rho_{0}}\left[  2-G_{0}\left(  1+\rho_{0}\right)
\right]  ,\nonumber\\
\beta &  =\frac{1}{2G_{0}}\left[  6-G_{0}\left(  1+2\rho_{0}\right)  \right]
,\quad\alpha=1,\nonumber\\
\delta &  =\frac{1}{3G_{0}\rho_{0}}\left[  6-G_{0}\left(  3+2\rho_{0}\right)
\right]  .
\end{align}
The metric collapses to Eq. (\ref{FRWb7}) so it is isotropic and
non-inflationary. As in the previous cases, since $\alpha=1,$ then the
potential $V$ is positive and behaves as $V\sim t^{-4},$ while $G\sim t^{2},$
i.e. $G$ is growing time function. If we set $\omega=0,$ then we obtain:%
\begin{align*}
\alpha &  =1,\quad\rho_{0}=\frac{1}{G_{0}}\left[  2-G_{0}\right]  ,\\
\beta &  =\frac{1}{2G_{0}}\left[  2+G_{0}\right]  ,\quad\delta=\frac{1}{3},
\end{align*}
in order to get, $\rho_{0}>0,$ we have the following restriction $G_{0}<2.$

\subsection{Scalar tensor model}

The effective stress-energy tensor takes the following form%
\begin{align}
T_{0}^{0}  &  =\frac{8\pi}{\phi}\rho-H\frac{\phi^{\prime}}{\phi}+\frac{\omega
}{2}\left(  \frac{\phi^{\prime}}{\phi}\right)  ^{2}+\Lambda\left(
\phi\right)  ,\\
T_{0}^{0}  &  =\left(  \frac{d^{\prime}}{d}-\frac{b^{\prime}}{b}\right)
\frac{\phi^{\prime}}{\phi}\\
T_{1}^{1}  &  =-\frac{8\pi}{\phi}p-\frac{\phi^{\prime}}{\phi}\left(
\frac{d^{\prime}}{d}+\frac{b^{\prime}}{b}\right)  -\frac{\omega}{2}\left(
\frac{\phi^{\prime}}{\phi}\right)  ^{2}-\frac{\phi^{\prime\prime}}{\phi
}+\Lambda\left(  \phi\right) \\
T_{2}^{2}  &  =-\frac{8\pi}{\phi}p-\frac{\phi^{\prime}}{\phi}\left(
\frac{a^{\prime}}{a}+\frac{b^{\prime}}{b}\right)  -\frac{\omega}{2}\left(
\frac{\phi^{\prime}}{\phi}\right)  ^{2}-\frac{\phi^{\prime\prime}}{\phi
}+\Lambda\left(  \phi\right) \\
T_{3}^{3}  &  =-\frac{8\pi}{\phi}p-\frac{\phi^{\prime}}{\phi}\left(
\frac{d^{\prime}}{d}+\frac{a^{\prime}}{a}\right)  -\frac{\omega}{2}\left(
\frac{\phi^{\prime}}{\phi}\right)  ^{2}-\frac{\phi^{\prime\prime}}{\phi
}+\Lambda\left(  \phi\right)
\end{align}
while the conservation equations are:%
\begin{align}
\left(  3+2\omega\right)  \left(  \frac{\phi^{\prime\prime}}{\phi}+H\frac
{\phi^{\prime}}{\phi}\right)  -2\left(  \Lambda-\phi\frac{d\Lambda}{d\phi
}\right)   &  =\frac{8\pi}{\phi}\left(  \rho-3p\right)  ,\\
\rho^{\prime}+\left(  \rho+p\right)  H  &  =0.
\end{align}

For the FE (\ref{JBD1}-\ref{JBD2}) and taking into account Eq. (\ref{JBDL}) we
find the next solution%

\begin{align}
a_{1}  &  =a_{2}=a_{3}=1,\quad q=0,\nonumber\\
\phi_{0}  &  =1,\quad\tilde{n}=-1-3\gamma,\nonumber\\
\rho_{0}  &  =-\frac{\left[  \omega\left(  3\gamma+1\right)  ^{2}+9\gamma
^{2}+12\gamma+1\right]  }{8\pi\left(  1+\gamma\right)  },\nonumber\\
\Lambda_{0}  &  =\frac{\left[  \omega\left(  \left(  1-\gamma\right)  \left(
3\gamma+1\right)  ^{2}\right)  +2\left(  3\gamma+1\right)  \right]  }{2\left(
1+\gamma\right)  }.
\end{align}
With these results, the metric collapses to Eq. (\ref{FRWb7}) and therefore it
is isotropic and non-inflationary. Therefore, $\rho_{0}=0,$ iff
\[
\gamma=-\frac{\left(  \omega\pm\sqrt{2\omega+3}+2\right)  }{3\omega+3}%
,\quad\omega\neq-1,\quad\omega\neq\frac{1}{2},
\]
while, $\Lambda_{0}=0,$ iff%
\[
\gamma=-\frac{1}{3},\,\gamma=\frac{\omega\pm\sqrt{2\omega\left(
2\omega+3\right)  }}{3\omega},\quad\omega\neq0,\,\omega\neq\frac{1}{2}.
\]
If we fix $\omega=3300,($\cite{BDparameter}$)$ then we get%
\begin{align*}
\rho_{0}  &  >0\quad\Longleftrightarrow\quad\gamma\in I_{1},\quad
I_{1}=\left(  -0.3416,-0.3252\right)  ,\\
\Lambda_{0}  &  <0\quad\Longleftrightarrow\quad\gamma\in I_{2},\quad
I_{2}=(-0.3334,-0.3333),
\end{align*}
in fact $\Lambda_{0}\geq0$ if $\gamma\notin I_{2}.$ Therefore this solution is
only valid $\forall\gamma\in I_{1},$ a small neighborhood of $\gamma_{c},$
$\mathcal{E}(\gamma_{c}=-\frac{1}{3})=I_{1},$ note that $I_{2}\subset I_{1},$
but if we take into account the current observations (\cite{SNIa}) which
suggest us that the cosmological constant must be possitive then we arrive to
the conclusion that the solution is valid $\forall\gamma\in I_{1}\backslash
I_{2}.$ If $\gamma>\gamma_{c},$ $\forall\gamma\in I_{1}$ then $\tilde{n}<0$
and therefore $G_{\mathrm{eff}}\thickapprox\phi^{-1},$ is a growing time
function, and $\Lambda$ behaves as a positive decreasing time function. For
example if $\gamma=-1/3,$ which means that $\tilde{n}=0,$ then
\[
\rho_{0}=\frac{3}{8\pi},\;\Lambda_{0}=0,\text{ }G_{\mathrm{eff}}=G_{0},
\]
compare with the perfect fluid solution and with classical model with TVC.
Note that $\tilde{n}>0,$ $\forall\gamma\in(-0.3416,\gamma_{c}),$ $\tilde{n}=0$
if $\gamma=\gamma_{c},$ and $\tilde{n}<0$, $\forall\gamma\in(\gamma
_{c},-0.3252).$

\section{Bianchi IX}

\subsection{The metric}

The Killings are (\cite{MC}):%
\begin{align*}
\xi_{1}  &  =\partial_{y},\\
\xi_{2}  &  =\cos y\partial_{x}-\cot x\sin y\partial_{y}+\frac{\sin y}{\sin
x}\partial_{z},\\
\xi_{3}  &  =-\sin y\partial_{x}-\cot x\cos y\partial_{y}+\frac{\cos y}{\sin
x}\partial_{z},
\end{align*}
therefore, $\left[  \xi_{1},\xi_{2}\right]  =\xi_{3},$ $\left[  \xi_{1}%
,\xi_{3}\right]  =-\xi_{2},$ $\left[  \xi_{2},\xi_{3}\right]  =\xi_{1}.$
Algebra brings us to get%
\begin{align}
ds^{2}  &  =-dt^{2}+\left(  a^{2}(t)\sin^{2}z+b^{2}(t)\cos^{2}z\right)
\mathrm{dx}^{2}+\nonumber\\
&  2\left(  b^{2}(t)-a^{2}(t)\right)  \sin z\sin x\cos z\mathrm{dxdy+}%
\nonumber\\
&  \left(  a^{2}(t)\sin^{2}x\cos^{2}z+b^{2}(t)\sin^{2}x\sin^{2}z\right.
+\nonumber\\
&  \left.  +d^{2}(t)\cos^{2}x\right)  \mathrm{dy}^{2}+2d^{2}(t)\cos
x\mathrm{dydz}+d^{2}(t)\mathrm{dz}^{2}. \label{m_BIX}%
\end{align}

The metric admits the following HVF%
\begin{equation}
\mathcal{H}=\left(  t+t_{0}\right)  \partial_{t}+\left(  1-a_{2}\right)
y\partial_{y}+\left(  1-a_{3}\right)  z\partial_{z}, \label{HO}%
\end{equation}
so the scale factors must behave as%
\[
a=a_{0}\left(  t+t_{0}\right)  ,\;b=b_{0}\left(  t+t_{0}\right)  ^{a_{2}%
},\;d=d_{0}\left(  t+t_{0}\right)  ^{a_{3}},
\]
and the restrictions on the constants $\left(  a_{i}\right)  _{i=1}^{3}%
\in\mathbb{R},$ are%
\begin{equation}
a_{1}=1,\qquad a_{2},a_{3}\in\mathbb{R}. \label{restrics}%
\end{equation}

For this model the FE read:%
\[
\frac{a^{\prime}}{a}\frac{b^{\prime}}{b}+\frac{a^{\prime}}{a}\frac{d^{\prime}%
}{d}+\frac{b^{\prime}}{b}\frac{d^{\prime}}{d}+\frac{1}{2}\left(  \frac
{1}{a^{2}}+\frac{1}{b^{2}}+\frac{1}{d^{2}}\right)
\]%
\begin{equation}
-\frac{1}{4}\left(  \frac{a^{2}}{b^{2}d^{2}}+\frac{b^{2}}{a^{2}d^{2}}%
+\frac{d^{2}}{b^{2}a^{2}}\right)  =8\pi GT_{1}^{1}, \label{fe91}%
\end{equation}%
\[
\frac{a^{\prime\prime}}{a}-\frac{b^{\prime\prime}}{b}+\frac{d^{\prime}}%
{d}\left(  \frac{a^{\prime}}{a}-\frac{b^{\prime}}{b}\right)
\]%
\begin{equation}
+\left(  \frac{1}{a^{2}}-\frac{1}{b^{2}}+\frac{a^{2}}{b^{2}d^{2}}-\frac{b^{2}%
}{a^{2}d^{2}}\right)  =0,
\end{equation}%
\[
\frac{b^{\prime\prime}}{b}+\frac{d^{\prime\prime}}{d}+\frac{b^{\prime}}%
{b}\frac{d^{\prime}}{d}+\frac{1}{2}\left(  \frac{1}{b^{2}}+\frac{1}{d^{2}%
}-\frac{1}{a^{2}}\right)
\]%
\begin{equation}
+\frac{1}{4}\left(  \frac{b^{2}}{a^{2}d^{2}}+\frac{d^{2}}{b^{2}a^{2}}%
-\frac{3a^{2}}{b^{2}d^{2}}\right)  =-8\pi GT_{2}^{2},
\end{equation}%
\[
\frac{a^{\prime\prime}}{a}+\frac{d^{\prime\prime}}{d}+\frac{a^{\prime}}%
{a}\frac{d^{\prime}}{d}+\frac{1}{2}\left(  \frac{1}{a^{2}}+\frac{1}{d^{2}%
}-\frac{1}{b^{2}}\right)
\]%
\begin{equation}
+\frac{1}{4}\left(  \frac{a^{2}}{b^{2}d^{2}}+\frac{d^{2}}{b^{2}a^{2}}%
-\frac{3b^{2}}{a^{2}d^{2}}\right)  =-8\pi GT_{3}^{3},
\end{equation}%
\[
\frac{b^{\prime\prime}}{b}-\frac{d^{\prime\prime}}{d}+\frac{a^{\prime}}%
{a}\frac{b^{\prime}}{b}-\frac{a^{\prime}}{a}\frac{d^{\prime}}{d}%
\]%
\begin{equation}
+\left(  \frac{1}{b^{2}}-\frac{1}{d^{2}}+\frac{b^{2}}{a^{2}d^{2}}-\frac{d^{2}%
}{b^{2}a^{2}}\right)  =0,
\end{equation}%
\[
\frac{a^{\prime\prime}}{a}+\frac{b^{\prime\prime}}{b}+\frac{a^{\prime}}%
{a}\frac{b^{\prime}}{b}+\frac{1}{2}\left(  \frac{1}{a^{2}}+\frac{1}{b^{2}%
}-\frac{1}{d^{2}}\right)
\]%
\begin{equation}
+\frac{1}{4}\left(  \frac{a^{2}}{b^{2}d^{2}}+\frac{b^{2}}{a^{2}d^{2}}%
-\frac{3d^{2}}{b^{2}a^{2}}\right)  =-8\pi GT_{4}^{4}, \label{fe96}%
\end{equation}

and the conservation equation(s).

\subsection{Classical solutions}

\subsubsection{Vacuum solution}

There is not SS solution for the vacuum model.

\subsubsection{Perfect fluid solution}

The FE\ \ are given by Eqs. (\ref{fe91}-\ref{fe96}) with the conservation
equation, $T_{i\,;j}^{j}=0.$ The stress-energy tensor is defined by Eq.
(\ref{eq00}), \ and the general behaviour for the solution is given by Eq.
(\ref{enerden}). We find the following solution
\begin{equation}
a_{1}=a_{2}=a_{3}=1,\;q=0,\;\omega=-1/3.
\end{equation}
In this way the metric collapses to this one:%
\begin{align}
ds^{2} &  =-dt^{2}+\nonumber\\
&  \left(  t+t_{0}\right)  ^{2}\left(  \mathrm{dx}^{2}+\mathrm{dy}%
^{2}+\mathrm{dz}^{2}\right)  +2\left(  t+t_{0}\right)  ^{2}\cos x\mathrm{dydz}%
\label{bixg4}%
\end{align}
and therefore there is another KVF, $\xi_{4}=\partial_{z}$. As in the previous
models this solution is isotropic, ($\mathcal{A},\mathcal{W}^{2})$ vanish and
it is only valid for the EoS $\omega=-1/3.$ The obtained solution is stable
from the dynamical point of view (\cite{ColeyDS}).

\subsubsection{Time varying constants framework}

In this framework the FE are given by Eqs. (\ref{fe91}-\ref{fe96}) with the
conservation equations (\ref{FE02}), where $\alpha=\left(  \omega+1\right)  h$
and $h=\left(  1+a_{2}+a_{3}\right)  $, and the stress-energy tensor is
defined by Eq. (\ref{eq00}). In this case the behaviour of the solution is
given by Eq. (\ref{sol-tvc-1}). Therefore we have found the next solution by
taking into account the obtained SS restrictions for the scale factors given
by Eq. (\ref{restrics})
\begin{align}
a_{1}  &  =a_{2}=a_{3}=1,\;q=0,\nonumber\\
G_{0}  &  =\frac{15}{48\pi\rho_{0}\left(  \omega+1\right)  },\quad
\forall\omega\in(-1,1],\nonumber\\
\Lambda &  =\frac{15}{4c^{2}}\left(  1-\frac{2}{3\left(  \omega+1\right)
}\right)  \left(  t+t_{0}\right)  ^{-2},
\end{align}
as above the metric collapses to Eq. (\ref{bixg4}), so it is isotropic and
non-inflationary. The behaviour of the \textquotedblleft
constants\textquotedblright\ is the following one:%
\begin{align*}
G  &  \thickapprox\left\{
\begin{array}
[c]{l}%
\text{decreasing }\forall\omega\in\lbrack-1,-1/3)\\
\text{constant if }\omega=-1/3\\
\text{growing }\forall\omega\in(-1/3,1]
\end{array}
\right. \\
\Lambda_{0}  &  \thickapprox\left\{
\begin{array}
[c]{l}%
\text{negative }\forall\omega\in\lbrack-1,-1/3)\\
\text{vanish if }\omega=-1/3\\
\text{positive }\forall\omega\in(-1/3,1]
\end{array}
\right.
\end{align*}
i.e. we have found the same behaviour as in the previous studied models.

\subsection{Scalar models}

\subsubsection{Scalar model}

The FE are given by Eqs. (\ref{fe91}-\ref{fe96}) while the stress-energy
tensor is defined by Eqs. (\ref{scalar}-\ref{defscal}). For this model the
conservation equation given by Eq. (\ref{CE-S1}). Hence, by taking into
account Eq. (\ref{RS1}) we find the following non-trivial solution
\begin{equation}
a_{1}=a_{2}=a_{3}=1,\qquad\alpha=\beta=\frac{5}{2},
\end{equation}
therefore the metric collapses to Eq. (\ref{bixg4}) and the solution is
isotropic and non-inflationary. \ As it is showed in (\cite{ColeyDS}) the
solution is stable from the DS point of view.

\subsubsection{Scalar model with G-var}

For this model the FE are given by Eqs. (\ref{fe91}-\ref{fe96}). The
stress-energy tensor is defined by Eq. (\ref{scalar}-\ref{defscal}) while the
conservation equation is defined by Eq. (\ref{G-var-Scal-1}). The solution
behaves as (\ref{RS2}). We find the following solution%
\begin{align}
a_{1}  &  =a_{2}=a_{3}=1,\quad q=0,\nonumber\\
\alpha &  =1,\quad\beta=1,\quad G_{0}=\frac{5}{2},
\end{align}
where, as it is observed, the metric collapses to Eq. (\ref{bixg4}) and
therefore the solution is isotropic and non-inflationary. Since, $\alpha=1,$
then we may say that $G\sim t^{2}$, i.e. it is a growing time function while
the potential is positive and decreasing, it behaves as, $V\sim t^{-4}.$

\subsubsection{Non-interacting scalar and matter model}

In this case, the FE for the model are described by Eqs. (\ref{fe91}%
-\ref{fe96}), and the stress-energy tensor is defined by Eqs. (\ref{eq00} and
\ref{scalar}-\ref{defscal}) while the conservation equations are defined by
Eqs. ($\left(  T^{m}\right)  _{i\,;j}^{j}=0=\left(  T^{\phi}\right)
_{i\,;j}^{j}$). The solution behaves as (\ref{RS3}). We find the following
solution%
\begin{align}
a_{1}  &  =a_{2}=a_{3}=1,\qquad q=0,\nonumber\\
\alpha &  =\beta,\qquad\rho_{0}=\frac{3}{2}\left(  \frac{5}{2}-\beta\right)
,\nonumber\\
\beta &  \in\left(  0,\frac{5}{2}\right)  ,\qquad\omega=-\frac{1}{3},
\end{align}
therefore the metric collapses to Eq. (\ref{bixg4}) and the solution is
non-inflationary and isotropic. As in the previous models, the solution is
only valid for $\omega=-\frac{1}{3},$ while the parameters $\left(
\alpha,\beta\right)  $ are only bounded.

\subsubsection{Non-interacting scalar and matter fields with G-var}

The model is described by the FE (\ref{fe91}-\ref{fe96}), with the
stress-energy tensor defined by Eqs. (\ref{eq00} and \ref{scalar}%
-\ref{defscal}) and the corresponding conservation equations given by Eqs
(\ref{NI1}-\ref{NI2}). The solution behaves as (\ref{RS4}), therefore we have
obtained the following solution
\begin{align}
a_{1}  &  =a_{2}=a_{3}=1,\qquad q=0,\nonumber\\
\alpha &  =1=\beta,\nonumber\\
\rho_{0}  &  =\frac{3}{4G_{0}}\left(  5-2G_{0}\right)  ,\quad\omega=-\frac
{1}{3},
\end{align}
so we have the following restriction, $G_{0}<\frac{5}{2},$ in order to get
$\rho_{0}>0.$ The metric collapses to Eq. (\ref{bixg4}) so the solution is
isotropic and non-inflationary and only valid for $\omega=-\frac{1}{3}.$ Note
that $\alpha=1,$ so $G$ is a growing time function, $G\sim t^{2},$ while the
potential is positive and behaves as a decreasing time function.

\subsubsection{Interacting scalar and matter model}

\paragraph{Approach 1}

The model is described by the FE (\ref{fe91}-\ref{fe96}) with the
stress-energy tensor is defined by Eqs. (\ref{eq00} and \ref{scalar}%
-\ref{defscal}) and conservation equation (\ref{Lima}). Since the behaviour of
the quantities are given by Eq. (\ref{RS3}), then we have found the following
solution%
\begin{align}
a_{1}  &  =a_{2}=a_{3}=1,\qquad q=0,\nonumber\\
\alpha &  =\frac{5}{2}-\rho_{0}\left(  \omega+1\right)  ,\quad\beta=\frac
{5}{2}+\frac{1}{2}\rho_{0}\left(  \omega-1\right)  ,
\end{align}
finding that the solution is valid $\forall\omega\in(-1,1],$ with the
restriction, $\rho_{0}<\frac{5}{2\left(  \omega+1\right)  }.$The metric
collapses to Eq. (\ref{bixg4}) which means that the solution is
non-inflationary and isotropic. For example if we set $\omega=0,$ then we get%
\[
\alpha=2\beta-\frac{5}{2},\quad\rho_{0}=5-2\beta,\quad\forall\beta\in\left(
\frac{5}{4},\frac{5}{2}\right)  .
\]

\paragraph{Approach 2}

The FE for this model are described by Eqs. (\ref{fe91}-\ref{fe96}) with the
stress-energy tensor is defined by Eqs. (\ref{eq00} and \ref{scalar}%
-\ref{defscal}) and the conservation equations are given by Eqs.
(\ref{Wett0}-\ref{Wett}). The behaviour of the quantities is given by Eq.
(\ref{RS3}), so we get the following solution
\begin{align}
a_{1}  &  =a_{2}=a_{3}=1,\qquad q=0,\nonumber\\
\alpha &  =\frac{5}{2}-\rho_{0}\left(  \omega+1\right)  ,\;\beta=\frac{5}%
{2}+\frac{1}{2}\rho_{0}\left(  \omega-1\right)  .\nonumber\\
\delta &  =-\frac{2\left(  1+3\omega\right)  }{\sqrt{10-4\rho_{0}\left(
\omega+1\right)  }},
\end{align}
finding the restriction, $\rho_{0}<\frac{5}{2\left(  \omega+1\right)  },$
$\forall\omega\in(-1/3,1],$ $\left(  \delta\leq0\right)  .$ With these
results, the metric collapses to Eq. (\ref{bixg4}) and therefore the solution
is isotropic and non-inflationary. For example if we set $\omega=0,$ then we
get%
\begin{align*}
\delta &  =-\frac{1}{\sqrt{\alpha}},\qquad\beta=\frac{5}{4}+\frac{\alpha}%
{2},\\
\rho_{0}  &  =\frac{5}{2}-\alpha,\qquad\forall\alpha\in\left(  0,\frac{5}%
{2}\right)  ,
\end{align*}
and therefore we have obtained the same result as in the approach 1, as it is expected.

\paragraph{Approach 3}

The FE (\ref{fe91}-\ref{fe96}) with the stress-energy tensor is defined by
Eqs. (\ref{eq00} and \ref{scalar}-\ref{defscal}) and conservation equations
given by Eqs.(\ref{Al1}-\ref{Al2}) admit the following solution, note that the
behaviour of the quantities is given by Eq. (\ref{RS3}),
\begin{align}
a_{1}  &  =a_{2}=a_{3}=1,\qquad q=0,\nonumber\\
\alpha &  =2-\frac{5}{2}\left(  \omega+1\right)  ,\quad\delta=\frac{1}%
{3}+\omega\nonumber\\
\beta &  =\frac{5}{2}+\frac{1}{2}\rho_{0}\left(  \omega-1\right)  ,
\end{align}
with $\rho_{0}=\rho_{0},$ finding the restriction, $\rho_{0}<\frac{5}{2\left(
\omega+1\right)  },$ $\forall\omega\in(-1/3,1],$ $\left(  \delta\geq0\right)
.$ Therefore the metric collapses to Eq. (\ref{bixg4}) which means that the
solution is non-inflationary and isotropic. If we set $\omega=0,$ then it is
obtained the following result:%
\begin{align*}
\delta &  =\frac{1}{3},\qquad\beta=\frac{5}{2}+\alpha,\\
\rho_{0}  &  =\frac{5}{2}-\alpha,\quad\forall\alpha\in\left(  0,\frac{5}%
{2}\right)  ,
\end{align*}
and therefore, once again we have shown that the three approaches are identical.

\subsubsection{Interacting scalar and matter fields with G-var}

The model is described by the FE (\ref{fe91}-\ref{fe96}) with the
stress-energy tensor defined by Eqs. (\ref{eq00} and \ref{scalar}%
-\ref{defscal}) while the conservation equations are given by Eqs.(\ref{ISG1}%
-\ref{ISG2}). In this case the behaviour of the quantities are given by Eq.
(\ref{RS6}), so we get the following solution with, $Q=\delta H\rho_{m},$%
\begin{align*}
a_{1}  &  =a_{2}=a_{3}=1,\qquad q=0,\quad\alpha=1,\\
\omega &  =\frac{1}{2G_{0}\rho_{0}}\left[  5-2G_{0}\left(  1+\rho_{0}\right)
\right]  ,\\
\beta &  =\frac{1}{4G_{0}}\left[  15-2G_{0}\left(  1+2\rho_{0}\right)
\right]  ,\\
\delta &  =\frac{1}{6G_{0}\rho_{0}}\left[  15-2G_{0}\left(  3+2\rho
_{0}\right)  \right]  .
\end{align*}
The metric collapses to Eq. (\ref{bixg4}), so the solution is isotropic and
non-inflationary. Since $\alpha=1,$ then $G$ is growing, it behaves as $G\sim
t^{2},$ while the potential behaves as a positive decreasing time function.
Setting $\omega=0,$ which may be interpreted as a model with interacting DM
and DE, then we obtain:%
\begin{align*}
\alpha &  =1,\quad\rho_{0}=\frac{1}{2G_{0}}\left[  5-2G_{0}\right]  ,\\
\beta &  =\frac{1}{4G_{0}}\left[  5+2G_{0}\right]  ,\quad\delta=\frac{1}{3},
\end{align*}
so we have the following restriction, $G_{0}<\frac{5}{2},$ in order to get
$\rho_{0}>0.$

\subsection{Scalar tensor model}

The effective stress-energy tensor takes the following form
\begin{align}
T_{1}^{1}  &  =\frac{8\pi}{\phi}\rho-H\frac{\phi^{\prime}}{\phi}+\frac{\omega
}{2}\left(  \frac{\phi^{\prime}}{\phi}\right)  ^{2}+\Lambda\left(
\phi\right)  ,\\
T_{2}^{2}  &  =-\frac{8\pi}{\phi}p-\frac{\phi^{\prime}}{\phi}\left(
\frac{d^{\prime}}{d}+\frac{b^{\prime}}{b}\right)  -\frac{\omega}{2}\left(
\frac{\phi^{\prime}}{\phi}\right)  ^{2}-\frac{\phi^{\prime\prime}}{\phi
}+\Lambda\left(  \phi\right)  ,\\
T_{3}^{3}  &  =-\frac{8\pi}{\phi}p-\frac{\phi^{\prime}}{\phi}\left(
\frac{d^{\prime}}{d}+\frac{a^{\prime}}{a}\right)  -\frac{\omega}{2}\left(
\frac{\phi^{\prime}}{\phi}\right)  ^{2}-\frac{\phi^{\prime\prime}}{\phi
}+\Lambda\left(  \phi\right)  ,\\
T_{4}^{4}  &  =-\frac{8\pi}{\phi}p-\frac{\phi^{\prime}}{\phi}\left(
\frac{a^{\prime}}{a}+\frac{b^{\prime}}{b}\right)  -\frac{\omega}{2}\left(
\frac{\phi^{\prime}}{\phi}\right)  ^{2}-\frac{\phi^{\prime\prime}}{\phi
}+\Lambda\left(  \phi\right)  ,
\end{align}
with the conservation equations%
\begin{align}
\left(  3+2\omega\left(  \phi\right)  \right)  \left(  \frac{\phi
^{\prime\prime}}{\phi}+H\frac{\phi^{\prime}}{\phi}\right)  -2\left(
\Lambda-\phi\frac{d\Lambda}{d\phi}\right)   &  =\frac{8\pi}{\phi}\left(
\rho-3p\right)  ,\\
\rho^{\prime}+\left(  \rho+p\right)  H  &  =0.
\end{align}

For the FE (\ref{JBD1}-\ref{JBD2}) and taking into account Eq. (\ref{JBDL}) we
find the next solution%

\begin{align}
a_{1}  &  =a_{2}=a_{3}=1,\qquad q=0\qquad\phi_{0}=1,\nonumber\\
\rho_{0}  &  =-\frac{\left[  2\omega\left(  3\gamma+1\right)  ^{2}%
+18\gamma^{2}+24\gamma+1\right]  }{16\pi\left(  1+\gamma\right)  },\nonumber\\
\Lambda_{0}  &  =\frac{\left[  2\omega\left(  \left(  1-\gamma\right)  \left(
3\gamma+1\right)  ^{2}\right)  +5\left(  3\gamma+1\right)  \right]  }{4\left(
1+\gamma\right)  },
\end{align}
therefore, the metric collapses to Eq. (\ref{bixg4}), so the solution is
isotropic and non-inflationary. $\rho_{0}=0,$ iff%
\[
\gamma=-\frac{\left(  2\omega\pm\sqrt{2\left(  5\omega+7\right)  }+4\right)
}{6\left(  \omega+1\right)  },\quad\omega\neq-1,
\]
and the cosmological constant vanish when, $\Lambda_{0}=0,$ iff%
\[
\gamma=-\frac{1}{3},\quad\gamma=\frac{2\omega\pm\sqrt{2\omega\left(
8\omega+15\right)  }}{6\omega},\quad\omega\neq0.
\]

If we fix $\omega=3300,$ (\cite{BDparameter}) then we get%
\begin{align*}
\rho_{0}  &  >0\quad\Longleftrightarrow\quad\gamma\in I_{1}\quad I_{1}=\left(
-0.3426,-0.3242\right)  ,\\
\Lambda_{0}  &  <0\quad\Longleftrightarrow\quad\gamma\in I_{2}\quad
I_{2}=(-0.3335,-0.3333),
\end{align*}
in fact $\Lambda_{0}\geq0$ if $\gamma\notin I_{2}.$ Note that $I_{2}\subset
I_{1}.$ Therefore this solution is only valid if $\gamma\in I_{1},$ i.e. for a
small neighborhood of $\gamma_{c},$ $\mathcal{E}(\gamma_{c}=-\frac{1}%
{3})=I_{1},$ note that $I_{2}\subset I_{1}.$ As in the above cases, if we
consider that $\Lambda_{0}$ must be possitive then we arrive to the conclusion
that the solution is valid $\forall\gamma\in I_{1}\backslash I_{2}.$ If
$\gamma>\gamma_{c},$ $\forall\gamma\in I_{1}$ then $\tilde{n}<0$ and therefore
$G_{\mathrm{eff}}\thickapprox\phi^{-1},$ is a growing time function, and
$\Lambda$ behaves as a positive decreasing time function. Setting
$\gamma=-1/3,$ then $\tilde{n}=0,$ and therefore%
\[
\rho_{0}=\frac{15}{32\pi},\quad\Lambda_{0}=0,\quad G_{eff}=G_{0},
\]
i.e. we obtain the classical perfect fluid solution.

\section{Conclusions}

In this paper we have studied how the constants $G$ and $\Lambda$ may vary in
different theoretical models as general relativity with a perfect fluid,
scalar cosmological models (\textquotedblleft quintessence\textquotedblright%
\ models) with and without interacting scalar and matter fields and a
scalar-tensor model with a dynamical $\Lambda$. We have applied the outlined
program to study three different geometries which generalize the FRW ones,
which are Bianchi \textrm{V}, \textrm{VII}$_{0}$ and \textrm{IX}, under the
self-similarity hypothesis. These geometries are in principle homogeneous but
anisotropic. We have put special emphasis in calculating exact power-law
solutions which allow us to compare the different models.

As we have shown, in all the studied cases, we arrive to very similar
conclusions. The first of them is that the solutions are isotropic
($\mathcal{A}$,$\mathcal{W}^{2}$ vanish) in spite of considering anisotropic
geometries, and noninflationary ($q=0$).

With regard to the classical solutions i.e. solutions obtained within the
general relativity (with a perfect fluid as matter model) and its TVC model,
we arrive to the conclusion that the obtained solutions are only valid for a
unique value of the EoS, $\omega=-1/3.$ In the Bianchi \textrm{V} model the
metric collapses to a metric with 6 KVF, therefore, as we already know, the FE
generalize the case FRW with negative curvature. In the Bianchi \textrm{VII}%
$_{0}$ model the metric collapses to the flat FRW and for the Bianchi
\textrm{IX} the obtained metric has only 4 KVF, nevertheless the FE generalize
the FRW with positive curvature. In the case of TVC models, we have arrived to
the conclusion that in all the cases the solutions are valid $\forall\omega
\in(-1,1]$. For $\omega=-1/3$ the solutions collapse to the standard solution
with $G=const.$ and $\Lambda=0.$ If we consider the current observations which
suggest $\Lambda_{0}>0,$ then we arrive to the conclusion that the solution is
valid only for $\forall\omega\in(-1/3,1]$ and that the gravitational
\textquotedblleft constant\textquotedblright, $G(t)\sim t^{3\omega+1}$,
behaves as a growing time function while $\Lambda\sim t^{-2}.$

With regard to the scalar cosmological models, we conclude that in all the
cases, the metrics collapse to the ones obtained for the case of a perfect
fluid model. All the obtained solutions are stable from the dynamical system
point of view and therefore relevant from the physical point of view. In the
case of the scalar models with $G-$var, we arrive to the conclusion that in
the cases $\mathrm{BVII}_{0}$ and $\mathrm{BIX,}$ $G$ behaves as a growing
time function, $G(t)\sim t^{2},$ and the potential, which mimics a dynamical
cosmological constant behaves as a positive decreasing function, $V\sim
t^{-4}.$ In the $\mathrm{BV}$ model, since it depends on more parameters like
$m,$ then the obtained solution is not so precise in such a way that we only
may say \ that if $\alpha\in(-1,\infty)\backslash\left\{  0\right\}  ,$ the
potential, $V(t)\sim\left(  t+t_{0}\right)  ^{-2\left(  \alpha+1\right)  },$
behaves as a decreasing time function, as a positive dynamical cosmological
constant, but $G$ is decreasing if $\alpha\in\left(  -1,0\right)  ,$ and it
behaves as a growing time function iff $\alpha>0$. In the non-interacting
case, we arrive to the conclusion that the obtained solutions are only valid
for a unique value of the EoS, $\omega=-1/3.$ In the non-interacting case with
$G-$var, the solutions are only valid for $\omega=-1/3,$ while $G$ and $V$
behave like in the case of the scalar model with $G-$var, so we have no new
information about their behavior. For the interesting cases of interacting
scalar and matter fields we arrive to the conclusion that for all the cases
the solutions are very similar. For the approaches 2 and 3 the solutions are
the same and only difference with respect to approach 1 is the range of
validity for $\omega,$ the parameter of the EoS$.$ The main difference with
respect to the non-interacting case is that the solutions are valid
$\forall\omega\in(-1/3,1]$ in approaches 2 and 3 while in approach 1
$\forall\omega\in(-1,1].$ This situation is more realistic from the physical
point of view$.$ For the critical value $\omega=-1/3$ the solutions collapse
to the non-interacting case in approaches 2 and 3. For the case of interacting
scalar and matter models with $G-$var we conclude that the solutions are valid
$\forall\omega\in(-1/3,1],$ while $G$ behaves as a growing time function
$G(t)\sim t^{2},$ and the potential behaves as a positive decreasing function,
$V\sim t^{-4},$ as in the above studied models.

As we have commented along the paper, the SSS may be considered as asymptotic
solutions for general solutions. In the same way we have pointed out that the
PF and the scalar solutions are stable from the DS point of view
(\cite{WE}-\cite{ColeyDS}). Since in the TVC scheme we have obtained the same
behaviour for the scale factors than in these solutions, we may
\textquotedblleft\emph{conjecture}\textquotedblright\ that the solutions
obtained in this framework are also stable, note that always we obtain
relationships like that, $G\rho_{m}\sim t^{-2}\sim G\rho_{\phi}.$ A simple
perturbation analysis may be carried out in order to prove this conjecture.

For the scalar-tensor theories we arrive in all the cases that the solution is
only valid for a small a small neighborhood of $\gamma_{c},$ $\mathcal{E}%
(\gamma_{c}=-\frac{1}{3})=I_{1}.$ For $\gamma_{c},$ then $\rho_{0}>0$ but
$\Lambda_{0}=0$ and $\tilde{n}=0,$ therefore $\phi=\phi_{0}$ and this means
that $G=G_{0}$ (i.e. constant), i.e. for this specific value, the solution is
the same than the one obtained in the classical model for a perfect fluid. In
the same way we may compare the JBD solution with the one obtained in the
classical situation when $G$ and $\Lambda$ vary. If $\gamma>\gamma_{c},$
$\forall\gamma\in I_{1}$ then $\tilde{n}<0$ and therefore $G_{\mathrm{eff}%
}\thickapprox\phi^{-1},$ is a growing time function, and $\Lambda$ behaves as
a positive decreasing time function. Note that $\tilde{n}>0,$ $\forall
\gamma\in\mathcal{E},$ $\gamma<\gamma_{c},$ $\tilde{n}=0$ if $\gamma
=\gamma_{c},$ and $\tilde{n}<0$ $\forall\gamma\in\mathcal{E},$ $\gamma
>\gamma_{c}.$ Once again, if we take into account the current observations,
then we arrive to the conclusion that $G$ behaves as a growing time function
$G(t)\sim t^{3\gamma-1},$ and the cosmological constant behaves as a positive
decreasing function, $\Lambda\sim t^{-2},$ as in the above studied models. We
have performed the same calculation considering the scalar-tensor theory
defined by
\[
S={\frac{1}{8\pi}}\int d^{4}x\sqrt{-g}\left\{  \frac{1}{2}\left[  \phi
R-\frac{\omega}{\phi}\phi_{,\alpha}\phi_{,}^{\alpha}-2U(\phi)\right]
+\mathcal{L}_{M}\right\}
\]
arriving to the same solutions that the obtained ones where, which is not
obvious as we have pointed out in (\cite{Tony2}).

\begin{acknowledgments}
I would like to thank A. Coley for value coments on the DS behaviour of the
solutions. In the same way I am very grateful to F. Navarro-L\'{e}rida for
reading and criticizing this work.
\end{acknowledgments}

\appendix

\section{Interacting scalar and matter models}

In this appendix we will prove the results stated in section \textrm{II}. We
shall study the admitted form for the potential, as well as, for the energy
density, in the SS framework. For this purpose we study the different DE
through the Lie group method (\cite{Lie}).

\subsection{Approach 1}

We may rewrite Eq. (\ref{Lima}) in an alternative way%
\begin{equation}
\rho_{m}^{\prime}+r\rho_{m}t^{-1}=-\phi^{\prime}\left(  \phi^{\prime\prime
}+h\phi^{\prime}t^{-1}+\frac{dV}{d\phi}\right)  , \label{IGA2}%
\end{equation}
where $r=\left(  \omega+1\right)  h,$ $H=ht^{-1}.$ Lie's method applied to Eq.
(\ref{IGA2}) brings us to get%
\begin{align}
t^{2}\xi_{\phi\phi}  &  =0,\label{l1}\\
-2t^{2}\xi_{t\phi}+2ht\xi_{\phi}+t^{2}\eta_{\phi\phi}  &  =0,\label{l2}\\
2t^{2}\eta_{t\phi}-t^{2}\xi_{tt}+3t^{2}\xi_{\phi}V_{\phi}+th\xi_{t}-h\xi &
=0, \label{l3}%
\end{align}%
\[
t^{2}\eta V_{\phi\phi}+t^{2}\eta_{tt}+4t\xi_{\phi}\left(  t\rho^{\prime}%
+r\rho\right)  +
\]%
\begin{equation}
t^{2}V_{\phi}\left(  2\xi_{t}-\eta_{\phi}\right)  +ht\eta_{t}=0, \label{l4}%
\end{equation}%
\[
t\rho^{\prime}\left(  3t\xi_{t}-2t\eta_{\phi}+r\xi\right)  +
\]%
\begin{equation}
r\rho\left(  3t\xi_{t}-2t\eta_{\phi}-\xi\right)  +t^{2}\xi\rho^{\prime\prime
}=0, \label{l5}%
\end{equation}%
\begin{equation}
\eta_{t}\left(  r\rho+t\rho^{\prime}\right)  =0. \label{l6}%
\end{equation}

Then, we may impose some symmetries in such a way that condition on the
integrability of the functions $V$ and $\rho$ are obtained. For example, the
symmetry $\left(  \xi=t,\eta=1\right)  ,$ which brings us to get the invariant
solution $\phi=\ln t,$ we get, from Eq. (\ref{l4})%
\[
V_{\phi\phi}+2V_{\phi}=0,\quad\Longleftrightarrow\quad V=\exp\left(
-2\phi\right)  ,
\]
and from Eq. (\ref{l5})
\begin{equation}
\rho^{\prime\prime}=-\left(  3+r\right)  \frac{\rho^{\prime}}{t}-2r\frac{\rho
}{t^{2}}\;\Longleftrightarrow\;\rho=C_{1}t^{-2}+C_{2}t^{-r},
\end{equation}
setting $C_{2}=0,$ we get%
\begin{equation}
\phi=\ln t,\quad V=\exp\left(  -2\phi\right)  ,\quad\rho=C_{1}t^{-2}.
\end{equation}

\subsection{Approach 2}

In this model the conservation equations read:%
\begin{align}
\rho_{m}^{\prime}+\left(  \omega+1\right)  \rho_{m}H  &  =-\phi^{\prime
}q^{\phi},\label{ana1}\\
\square\phi+\frac{dV}{d\phi}  &  =q^{\phi}, \label{ana2}%
\end{align}
If we assume such functional relationship, $q^{\phi}=\beta\rho_{m},$ then we
obtain for Eq. (\ref{ana2}) the following system of PDE%
\begin{align}
t^{2}\xi_{\phi\phi}  &  =0,\label{w1}\\
-2t^{2}\xi_{t\phi}+2ht\xi_{\phi}+t^{2}\eta_{\phi\phi}  &  =0,\label{w2}\\
3t^{2}\xi_{\phi}\left(  V_{\phi}-\beta\rho\right)  +th\xi_{t}-h\xi+2t^{2}%
\eta_{t\phi}-t^{2}\xi_{tt}  &  =0, \label{w3}%
\end{align}%
\[
t^{2}\eta V_{\phi\phi}+t^{2}\eta_{tt}+2t^{2}\xi_{t}\left(  V_{\phi}-\beta
\rho\right)  -
\]%
\begin{equation}
t^{2}\eta_{\phi}\left(  V_{\phi}-\beta\rho\right)  +ht\eta_{t}-\beta t^{2}%
\xi\rho^{\prime}=0. \label{w4}%
\end{equation}
The symmetry $\left(  \xi=t,\eta=1\right)  \Longrightarrow\phi=\ln t,$ brings
us to obtain the following restriction, from Eq. (\ref{w4}) we get%
\begin{align*}
V_{\phi\phi}+2V_{\phi}  &  =\beta\left(  t\rho^{\prime}+2\rho\right)  ,\\
V  &  =\exp\left(  -2\phi\right)  =\beta\rho\thickapprox t^{-2},
\end{align*}

While Eq. (\ref{ana1}) admits the following solutions%
\[
\phi=-\frac{1}{\beta}\left(  r\ln t-\ln\rho\right)  +C_{1},
\]
in an alternative way%
\[
\rho=\exp\left[  -\beta\phi-r\ln t+\beta C_{1}\right]  =\rho_{0}\frac
{1}{t^{r+\beta}},
\]
$r+\beta=2.$ Therefore we have obtained:%
\begin{equation}
\phi=\ln t,\qquad V=\exp\left(  -2\phi\right)  ,\qquad\rho=\rho_{0}t^{-2}.
\end{equation}

\subsection{Approach 3}

The conservation equations read%

\begin{align}
\rho_{m}^{\prime}+\left(  \omega+1\right)  \rho_{m}H  &  =\delta H\rho
_{m},\label{peta1}\\
\phi^{\prime}\left(  \square\phi+\frac{dV}{d\phi}\right)   &  =-\delta
H\rho_{m}, \label{peta2}%
\end{align}

From Eq. (\ref{peta1}) we get
\[
\rho_{m}=\rho_{0}t^{h\left(  \delta-\left(  \omega+1\right)  \right)  },
\]
and therefore Eq. (\ref{peta2}) yields
\begin{equation}
\phi^{\prime}\left(  \phi^{\prime\prime}+h\phi^{\prime}t^{-1}+\frac{dV}{d\phi
}\right)  =-At^{a}, \label{he}%
\end{equation}
with $a=h\left(  \delta-\left(  \omega+1\right)  \right)  -1,$ $A=\delta
h\rho_{0}.$ We have the next system of PDE for Eq. (\ref{he})%
\begin{align}
\xi_{\phi\phi}t^{2}  &  =0,\label{a1}\\
-2t^{2}\xi_{t\phi}+2ht\xi_{\phi}+t^{2}\eta_{\phi\phi}  &  =0,\label{a2}\\
3t^{2}\xi_{\phi}V_{\phi}+th\xi_{t}-h\xi+2t^{2}\eta_{t\phi}-t^{2}\xi_{tt}  &
=0,\label{a3}\\
3At^{2+a}\xi_{t}-2At^{2+a}\eta_{\phi}+aAt^{1+a}\xi &  =0,\label{a4}\\
At^{2+a}\eta_{t}  &  =0. \label{a5}%
\end{align}%
\[
t^{2}\eta V_{\phi\phi}+t^{2}\eta_{tt}+4At^{2+a}\xi_{\phi}V_{\phi}-
\]%
\begin{equation}
t^{2}\eta_{\phi}V_{\phi}+ht\eta_{t}+2t^{2}\xi_{t}V_{\phi}=0, \label{a6}%
\end{equation}

The symmetry $\left(  \xi=t,\eta=1\right)  ,$ bring us to get from Eq.
(\ref{a6})%
\begin{align*}
t^{2}\eta V_{\phi\phi}+2t^{2}\xi_{t}V_{\phi}  &  =0,\\
3At^{2+a}\xi_{t}+aAt^{1+a}\xi &  =0,
\end{align*}
simplifying%
\[
V_{\phi\phi}+2\alpha V_{\phi}=0\qquad V=V_{0}\exp(-2\alpha\phi),
\]
where $a=-3.$ Therefore we get:%
\begin{equation}
\rho_{m}=\rho_{0}t^{-2},\qquad\phi=\ln t,\qquad V=\exp\left(  -2\phi\right)  .
\end{equation}

\subsection{A generalization}

In this model we try to generalize the previous approaches. We consider a
function $Q=\delta Hf(t),$ $f(t)$ is an unknown function. Therefore applying
the same procedure we would like to determine the admitted form for $f$ in
order to obtain a SSS. For this purpose we consider the following system of DE%

\begin{align}
\rho_{m}^{\prime}+\left(  \omega+1\right)  \rho_{m}H  &  =\delta
Hf(t),\label{alisa1}\\
\phi^{\prime}\left(  \square\phi+\frac{dV}{d\phi}\right)   &  =-\delta Hf(t),
\label{alisa2}%
\end{align}
From Eq. (\ref{alisa1}) we get%
\begin{equation}
\rho=\left(  \delta h\int f(t)t^{r-1}dt+C_{1}\right)  t^{-r}. \label{ma}%
\end{equation}
where $r=h\left(  \omega+1\right)  ,$ and $H=ht^{-1}.$ In the same way we may
analyze Eq. (\ref{alisa2}) through the Lie group method%
\begin{equation}
\phi^{\prime}\left(  \phi^{\prime\prime}+h\phi^{\prime}t^{-1}+\frac{dV}{d\phi
}\right)  =-\delta ht^{-1}f(t),
\end{equation}
getting the following system of PDE:%
\begin{align}
\xi_{\phi\phi}t^{2}  &  =0,\label{m1}\\
-2t^{2}\xi_{t\phi}+2ht\xi_{\phi}+t^{2}\eta_{\phi\phi}  &  =0,\label{m2}\\
3t^{2}\xi_{\phi}V_{\phi}+th\xi_{t}-h\xi+2t^{2}\eta_{t\phi}-t^{2}\xi_{tt}  &
=0,\label{m3}\\
3\delta htf(t)\xi_{t}-2\delta htf(t)\eta_{\phi}+\delta h\left(  tf^{\prime
}(t)-f\right)  \xi &  =0,\label{m4}\\
\delta htf(t)\eta_{t}  &  =0, \label{m5}%
\end{align}%
\[
t^{2}\eta V_{\phi\phi}+t^{2}\eta_{tt}+4\delta htf(t)\xi_{\phi}V_{\phi}-
\]%
\begin{equation}
t^{2}\eta_{\phi}V_{\phi}+ht\eta_{t}+2t^{2}\xi_{t}V_{\phi}=0. \label{m6}%
\end{equation}

The symmetry $\left(  \xi=t,\eta=1\right)  ,$bring us to get from Eq.
(\ref{m6})
\begin{align*}
V_{\phi\phi}+2V_{\phi}  &  =0\qquad V=V_{0}\exp(-2\phi),\\
tf^{\prime}(t)+2f  &  =0\qquad f=f_{0}t^{-2},
\end{align*}
then Eq. (\ref{ma}) gives%
\begin{equation}
\rho=\left(  \delta h\int t^{r-3}dt+C_{1}\right)  t^{-r},
\end{equation}
and therefore%
\[
\rho=C_{1}t^{-2}+C_{2}t^{-r}%
\]

Therefore, setting $C_{2}=0,$ we have obtained the following results:%
\begin{equation}
\rho_{m}=\rho_{0}t^{-2},\;\phi=\ln t,\;V=\exp\left(  -2\phi\right)
,\;f=f_{0}t^{-2}.
\end{equation}

As we can see we can consider any function $f$ that behaves as $f=f_{0}%
t^{-2}.$ This fact justifies the employ of $f=\rho_{m},$ as in the previous
approaches, but we can also consider other forms for $f$ as for example,
$f=\rho_{\phi}\thickapprox t^{-2},$ or $f=\rho_{m}+\rho_{\phi}\thickapprox
t^{-2},$ etc.

\section{Matter and scalar fields with $G-$varying}

In this appendix we will study the equation%
\[
\rho_{m}^{\prime}+\phi^{\prime\prime}\phi^{\prime}+\frac{dV}{d\phi}%
\phi^{\prime}+H\left(  \phi^{\prime2}+\left(  \rho_{m}+p_{m}\right)  \right)
\]%
\begin{equation}
=-\frac{G^{\prime}}{G}\left(  \rho_{m}+\frac{1}{2}\phi^{\prime2}+V\right)  ,
\label{Gv3}%
\end{equation}
that we may study as follows. In the first case and in analogy with the
previous appendix we will study the whole equation i.e. Eq. (\ref{Gv3})
without spliting it while in the second case we will split it for a suitable
coupling function.

\subsection{Case 1}

Eq. (\ref{Gv3}) may be written in the following form%
\[
\left(  \phi^{\prime\prime}+\left(  ht^{-1}+\frac{G^{\prime}}{2G}\right)
\phi^{\prime}+\frac{dV}{d\phi}\right)  \phi^{\prime}%
\]%
\[
+\frac{G^{\prime}}{G}V+\rho_{m}^{\prime}+\left(  rt^{-1}+\frac{G^{\prime}}%
{G}\right)  \rho_{m}=0,
\]
where $r=\left(  1+\gamma\right)  h$ and $H=ht^{-1}.$ The standard procedure
brings us to get the following system of PDE%
\begin{align}
t^{2}G^{2}\xi_{\phi\phi}  &  =0,\label{tif1}\\
2t^{2}G^{2}\eta_{\phi\phi}-4t^{2}G^{2}\xi_{t\phi}+2tG\xi_{\phi}\left(
2hG+tG^{\prime}\right)   &  =0, \label{tif3}%
\end{align}%
\[
6tG^{2}\xi_{\phi}\left(  \rho+tV_{\phi}\right)  +4t^{2}G^{2}\eta_{t\phi
}-2t^{2}G^{2}\xi_{tt}+
\]%
\begin{equation}
\left(  -2hG^{2}-t^{2}G^{\prime2}+t^{2}GG^{\prime\prime}\right)  \xi+tG\left(
2hG+tG^{\prime}\right)  \xi_{t}=0, \label{tif3a}%
\end{equation}%
\begin{align*}
&  4t^{2}G^{2}V_{\phi}\xi_{t}-2t^{2}G^{2}V_{\phi}\eta_{\phi}+\\
&  8tG\left(  tG^{\prime}\left(  V+\rho\right)  +r\rho G+tG\rho^{\prime
}\right)  \xi_{\phi}+
\end{align*}%
\begin{equation}
2t^{2}G^{2}\eta_{tt}+2t^{2}G^{2}V_{\phi\phi}\eta+tG\left(  2hG+tG^{\prime
}\right)  \eta_{t}=0, \label{tif4}%
\end{equation}

\begin{align*}
&  2t^{2}GG^{\prime}V_{\phi}\eta+6tG\left(  tG^{\prime}V+r\rho G+tG^{\prime
}\rho+tG\rho^{\prime}\right)  \xi_{t}-\\
&  4tG\left(  tG^{\prime}V+r\rho G+tG^{\prime}\rho+tG\rho^{\prime}\right)
\eta_{\phi}+
\end{align*}%
\[
2t^{2}G\left(  G^{\prime\prime}V+G^{\prime}\rho^{\prime}+t^{-1}Gr\rho^{\prime
}+G^{\prime\prime}\rho\right)  \xi
\]%
\begin{equation}
+2t^{2}G\left(  G\rho^{\prime\prime}-r\rho Gt^{-2}-G^{-1}G^{\prime2}\left(
V+\rho\right)  \right)  \xi=0, \label{tif5}%
\end{equation}%
\begin{equation}
2t^{2}G^{2}\left(  \rho+G^{-1}G^{\prime}V+t^{-1}r\rho+G^{\prime}\rho\right)
\eta_{t}=0. \label{tif6}%
\end{equation}

The symmetry $\left(  \xi=t,\quad\eta=\alpha\phi\right)  \Longrightarrow
\phi=t^{\alpha},$ bring us to obtain the following restriction on the
functions $G$, $V$ and $\rho$. From (\ref{tif3a}) we obtain%
\[
G^{\prime\prime}=\frac{G^{\prime2}}{G}-\frac{G^{\prime}}{t},\qquad
\Longrightarrow\qquad G=G_{0}t^{g}.
\]

Now from Eq. (\ref{tif4}) we get%
\[
V_{\phi\phi}=\left(  \frac{\alpha-2}{\alpha}\right)  \frac{V_{\phi}}{\phi
},\qquad\Longrightarrow\qquad V\left(  \phi\right)  =V_{0}\phi^{^{\frac
{2}{\alpha}\left(  \alpha-1\right)  }}.
\]

From Eq. (\ref{tif5}) we obtain since $\phi=t^{\alpha}$ and $V\left(
\phi\right)  =V_{0}\phi^{^{\frac{2}{\alpha}\left(  \alpha-1\right)  }}%
=V_{0}t^{2\left(  \alpha-1\right)  },$ and $G=G_{0}t^{g}$ then we get%
\[
\rho^{\prime\prime}=-c_{1}\frac{\rho^{\prime}}{t}-c_{2}\frac{\rho}{t^{2}%
}-c_{3}t^{2(\alpha-2)},
\]
finding that a solution is:%
\begin{equation}
\rho_{m}=\rho_{0}\left(  t+t_{0}\right)  ^{2\left(  \alpha-1\right)  }.
\end{equation}

Therefore we get (after redefining the constants)%
\begin{align}
\phi &  =\phi_{0}\left(  t+t_{0}\right)  ^{-\alpha},\;V(t)=\beta\left(
t+t_{0}\right)  ^{-2\left(  \alpha+1\right)  },\nonumber\\
G  &  =G_{0}\left(  t+t_{0}\right)  ^{g},\;\rho_{m}=\rho_{0}\left(
t+t_{0}\right)  ^{-2\left(  \alpha+1\right)  },
\end{align}
with $g=2\alpha,$ since $G\rho\thickapprox t^{-2},$ and $p_{m}=\omega\rho
_{m}.$

\subsection{Case 2}

In this case we may split Eq. (\ref{Gv3}) as follows%
\begin{align}
\rho_{m}^{\prime}+\left(  \rho_{m}+p_{m}\right)  H+\frac{G^{\prime}}{G}%
\rho_{m}  &  =\delta H\rho_{m},\label{pet1}\\
\phi^{\prime}\left(  \phi^{\prime\prime}+\phi^{\prime}H+\frac{dV}{d\phi
}\right)  +\frac{G^{\prime}}{G}\rho_{\phi}  &  =-\delta H\rho_{m},
\label{pet2}%
\end{align}
in such a way that if we study both equations through the LG method then we get.

For Eq. (\ref{pet2}) we get:%
\begin{align}
t^{2}G^{2}\xi_{\phi\phi}  &  =0,\label{imf1}\\
2t^{2}G^{2}\eta_{\phi\phi}-4t^{2}G^{2}\xi_{t\phi}+2tG\xi_{\phi}\left(
2hG+tG^{\prime}\right)   &  =0, \label{imf2}%
\end{align}%
\[
4t^{2}G^{2}\eta_{t\phi}-2t^{2}G^{2}\xi_{tt}+6t^{2}G^{2}V_{\phi}\xi_{\phi}+
\]%
\begin{equation}
\left(  -2hG^{2}-t^{2}G^{\prime2}+t^{2}GG^{\prime\prime}\right)  \xi+tG\left(
2hG+tG^{\prime}\right)  \xi_{t}=0, \label{imf3}%
\end{equation}%
\[
8tG\left(  tG^{\prime}V_{\phi}+\delta h\rho\right)  \xi_{\phi}+2t^{2}G^{2}%
\eta_{tt}+tG\left(  2hG+tG^{\prime}\right)  \eta_{t}%
\]%
\begin{equation}
+2t^{2}G^{2}V_{\phi\phi}\eta-2t^{2}G^{2}V_{\phi}\eta_{\phi}+4t^{2}G^{2}%
V_{\phi}\xi_{t}=0, \label{ifm4}%
\end{equation}%
\begin{align*}
&  2t^{2}GG^{\prime}V_{\phi\phi}\eta+6tG\left(  tG^{\prime}V_{\phi}+G\delta
h\rho\right)  \xi_{t}\\
&  -4tG\left(  tG^{\prime}V_{\phi}+G\delta h\rho\right)  \eta_{\phi}+
\end{align*}%
\begin{align}
2\left(  \delta hG^{2}\left(  t\rho^{\prime}-\rho\right)  +2t^{2}V_{\phi
}\left(  GG^{\prime\prime}-G^{\prime2}\right)  \right)  \xi &  =0,\label{ifm5}%
\\
2tG\left(  tG^{\prime}V_{\phi}+G\delta h\rho\right)  \eta_{t}  &  =0.
\label{ifm6}%
\end{align}

The symmetry $\left(  \xi=t,\text{ }\eta=\alpha\phi\right)  \Longrightarrow
\phi=t^{\alpha},$ bring us to get from Eq. (\ref{imf3})%
\[
G^{\prime\prime}=\frac{G^{\prime2}}{G}-\frac{G^{\prime}}{t}\Longrightarrow
G=G_{0}t^{g}.
\]
From Eq. (\ref{ifm4})%
\[
V_{\phi\phi}=\frac{\left(  \alpha-2\right)  }{\alpha}\frac{V_{\phi}}{\phi
}\quad\Longrightarrow\quad V=V_{0}\phi^{\frac{2\left(  \alpha-1\right)
}{\alpha}}.
\]
Now, from Eq. (\ref{ifm5}) we obtain an ODE for the energy density
\begin{equation}
2\left(  1-\alpha\right)  \rho+\rho^{\prime}t=0,\quad\Longrightarrow\quad
\rho=\rho_{0}t^{2\left(  \alpha-1\right)  }.
\end{equation}
Note that for Eq. (\ref{pet1}) we get a direct integration since%
\[
\rho_{m}^{\prime}+\left(  1+\gamma-\delta\right)  H\rho_{m}+\frac{G^{\prime}%
}{G}\rho_{m}=0,
\]
then $\rho_{m}G=Kt^{-r},$ where $r=\left(  1+\gamma-\delta\right)  .$

Therefore we have arrived to the same conclusion as in the above case i.e.:%
\begin{align}
\phi &  =\phi_{0}\left(  t+t_{0}\right)  ^{-\alpha},\;V(t)=\beta\left(
t+t_{0}\right)  ^{-2\left(  \alpha+1\right)  },\nonumber\\
G  &  =G_{0}\left(  t+t_{0}\right)  ^{g},\;\rho_{m}=\rho_{0}\left(
t+t_{0}\right)  ^{-2\left(  \alpha+1\right)  },
\end{align}
with $g=2\alpha,$ since $G\rho\thickapprox t^{-2}.$


\begin{thebibliography}{99}                                                                                               %


\bibitem {Iranies}Adabi, F. et al, arXiv:1105.1008. Sheykhi, A. and Setare,
M.R., IJTP \textbf{49}, 2777 (2010).

\bibitem {Tony1}Belinch\'{o}n, J.A.: ArXiv:1109.2880.

\bibitem {Tony2}Belinch\'{o}n, J.A.: ArXiv:1109.2877.

\bibitem {Alan1}Billyard, A.P and Coley, A. Phys Rev D\textbf{61}, 083503 (2000).

\bibitem {KB}Bronnikov, K.A et al. Class. Quantum Grav. \textbf{21}, 3389-3403 (2004).

\bibitem {Carr}Carr, B. J. and Coley, A., Class. Quantum Grav. 16, R31 (1999).

\bibitem {ColeyDS}Coley, A.: \textquotedblleft Dynamical Systems and
Cosmology\textquotedblright. Kluwer Academic Publishers (2003).

\bibitem {Ellis}Ellis, G.F.R. and Madsen, M.S., Class. Quantum Grav.
\textbf{8}, 667 (1991).

\bibitem {H}Harko T and Mak M K.: Int. J. Mod. Phys. D\textbf{11}, 1171 (2002).

\bibitem {HW}Hsu, L. and Wainwright, J. Class. Quantum Grav, \textbf{3}, 1105-24,(1986).

\bibitem {Lie}Ibragimov, N. H. \textquotedblleft Elementary Lie Group Analysis
and Ordinary Differential Equations\textquotedblright. Jonh Wiley \& Sons,
(1999). Bluman, G.W \ and Anco, S.C. \textquotedblleft Symmetry and Integral
Methods for Differential Equations\textquotedblright. Springer-Verlang (2002).

\bibitem {Griego}Jamil, M. et al, Phys.Lett.B\textbf{679}, 172 (2009). Lu, J.
et al, JCAP 1003:031 (2010).

\bibitem {KM}Kitada, Y. and Maeda, Kei-ichi, Class. Quantum Grav. \textbf{10}
703 (1993).

\bibitem {Krori}Krori, K.D. et al: Gen. Rel. Grav. \textbf{32}, 1439 (2000).

\bibitem {Lau}Lau, Y-K.: Aust. J. Phys. \textbf{38}, 547 (1985).

\bibitem {Lima}Maia, J.M.F., Lima, J.A.S., Phys.Rev. D\textbf{65}, 083513 (2002).

\bibitem {Nilsson}Nilsson, U.S. et al. Astro. Jour., \textbf{521}, L1--L3 (1999).

\bibitem {Pavon1}Pav\'{o}n, D. and Wang, B., Gen. Rel. Grav. \textbf{41}, 1 (2009).

\bibitem {SNIa}Perlmutter, S. et al., Astrophys. J. \textbf{483}, 565 (1997).
Nature \textbf{391}, 51 (1998,). Astrophys. J. \textbf{517}, 565 (1999).

\bibitem {Prigo}Prigogine, I. et al, Gen. Relativ. Gravit. \textbf{21}, 767 (1989).

\bibitem {MC}Stephani, H. et al., \textquotedblleft Exact Solutions of
Einstein's Field Equations\textquotedblright\ (2nd edn.). Cambridge University
Press (2003).

\bibitem {BDparameter}Susperregi, M. and Mazumdar, A. Phys. Rev. D
\textbf{58}, 083512 (1998).

\bibitem {Wainwrit}Wainwright, J. : \textquotedblleft Self-Similar Solutions
of Einstein's Equations\textquotedblright. Published in Galaxies, Axisymmetric
Systems \& Relativity. Ed M.A.H MacCallum. Cambridge University Press (1985).

\bibitem {WE}Wainwright, J., Ellis, G.F.R., \textquotedblleft Dynamical
Systems in Cosmology\textquotedblright. Cambridge University Press (1997).

\bibitem {W}Wainwright, J., Hancock M.J. and Uggla C.: Class. Quantum Grav.
\textbf{16}, 2577 (1999)

\bibitem {Werterich}Wetterich, C. Astron. Astrophys \textbf{301}, 321-328 (1995).

\bibitem {wett2}Wetterich, C., Nucl. Phys. B\textbf{302}, 668 (1988).

\bibitem {will}Will, C.M., \textquotedblleft Theory and experiments in
gravitational physics\textquotedblright. Cambridge University Press (1993).
\end{thebibliography}
\end{document}